\documentclass[11pt,a4paper]{article}
\PassOptionsToPackage{
  colorlinks=true,
  linkcolor=blue,
  citecolor=blue,
  urlcolor=blue
}{hyperref}

\usepackage[utf8]{inputenc}
\usepackage[T1]{fontenc}
\usepackage[english]{babel}
\usepackage{amsmath,amssymb,amsthm}
\usepackage{graphicx}
\usepackage{booktabs}
\usepackage{siunitx}
\usepackage{adjustbox}
\usepackage{makecell}
\usepackage{tablefootnote}
\usepackage{float}
\usepackage{soul}
\usepackage{enumerate}
\usepackage{comment}
\usepackage{placeins}
\usepackage{threeparttable}
\usepackage{multirow}
\usepackage{xcolor}

\usepackage[
  authordate,           
  backend=biber,
  natbib=true,
  sorting=nyt,
  isbn=false,
  url=false
]{biblatex-chicago}
\usepackage{hyperref}
\usepackage{orcidlink}

\usepackage[margin=2.5cm]{geometry}

\title{Beyond the Trade-off Curve: Multivariate and Advanced Risk-Utility Maps for Evaluating Anonymized Data}

\author{
  Oscar Thees\orcidlink{0009-0001-9378-4988}$^*$
  \and
  Roman Müller\orcidlink{0009-0007-2142-3896}$^*$
  \and
  Matthias Templ\orcidlink{0000-0002-8638-5276}$^*$
}

\date{
  \today \\[0.3em]
  \small $^{*}$University of Applied Sciences and Arts Northwestern Switzerland \\
  \small Corresponding author: Matthias Templ
}
\date{}

\begin{document}

\maketitle

\begin{abstract}
Anonymizing microdata requires balancing disclosure risk reduction with the preservation of data utility. Traditional evaluations often rely on single measures or two-dimensional risk-utility (R-U) maps, but real-world assessments involve multiple, often correlated, indicators of both risk and utility — a fundamentally multivariate problem that pairwise comparisons fail to capture both efficiently and completely. We systematically compare six visualization approaches for simultaneous evaluation of multiple risk and utility measures: heatmaps, dot plots, composite scatterplots, parallel coordinate plots, radial profile charts, and principal component analysis (PCA)-based biplots. We introduce blockwise PCA for composite scatterplots and joint PCA for biplots that simultaneously reveal method performance and measure interrelationships, and apply systematic Pareto-optimal method identification across all approaches where applicable, with dominance assessed in the original composite score space. Our comparison shows that no single approach dominates across all criteria: PCA biplots best reveal multivariate structure, while composite scatterplots offer intuitive summaries accessible to broader audiences. Combining complementary visualizations provides the most complete basis for evaluating the risk-utility trade-off.
\end{abstract}


\section{Introduction}\label{sec:intro}

In microdata anonymization, modifications to the original dataset, such as suppression, generalization, perturbation, or synthetic data generation, are essential to reduce disclosure risk \parencite{2012_Hundepool, 2017_Templ}. However, these modifications inevitably result in a loss of information, potentially limiting the analytical value of the data. Effective anonymization therefore requires balancing the confidentiality gained from lowering disclosure risk with the preservation of data utility. A possibility for evaluating this trade-off visually is the risk–utility (R-U) confidentiality map \parencite{2001_Duncan}. R–U maps depict anonymized datasets according to their measured disclosure risk and data utility, facilitating the systematic comparison of alternative anonymization approaches. By anonymization approaches, we refer to the different strategies by which datasets can be anonymized. These may involve method-specific variations, such as adjusting the level of noise in noise addition \parencite{Brand_2002_Microdata} or comparisons between methods – e.g., the effect of microaggregation \parencite{Defays_1998_Masking} and post randomization \parencite{Gouweleeuw_1998_Post} on disclosure risk. They can also include differences in synthetic data, either through employing distinct data generators and/or by producing multiple datasets with the same generator. Although the visualization methods presented in this paper are illustrated using synthetic data examples, the proposed visualizations can also be applied to traditional anonymization methods. The prerequisite is that the utility and risk measures used for comparison are appropriate and comparable across all applied anonymization approaches.

In practice, it is common to report one or more disclosure risk indicators and utility measures separately, often presented as summary statistics in tables or text, or visualized through simplified two-dimensional R–U maps \parencite[e.g.,][]{Muralidhar_2006_Data, Templ_2008_Robust, Hornby_2021_Identification, Little_2022_Comparing, Little_2025_Synthetica}. Although such representations, especially R-U maps, aid in interpretation, displaying several risk and utility measures quickly becomes cumbersome. One remedy is to use faceting and arrange multiple R–U maps as \textit{small multiples}: by enforcing comparisons of change, of differences between objects, small multiples are often the best solution for a wide range of presentation problems \parencite[p.~67-68]{Tufte_1990_Envisioning}. However, as the number of measures grows and thereby the number of panels needed for their display, this approach becomes more fragmented and inefficient \parencite{Hosseinpour_2025_Examining}. For example, with five distinct risk measures and five utility measures, 25 different R–U plots would be required to examine all pairwise relationships, complicating overall interpretation.

This challenge stems from the inherently multivariate nature of the evaluation problem: different risk and utility measures reflect distinct, often uncorrelated, aspects of data protection and analytical validity. Presenting them separately obscures potential interactions and trade-offs across dimensions. 
    While \textcite{2022_Dankar_Multi} explicitly acknowledge the multidimensional nature of utility by highlighting the large number of available utility metrics and the absence of general guidelines or standardised thresholds for their interpretation, they also propose dimension-reduction approaches, such as principal component analysis, to summarize multiple utility measures \parencite{2022_Dankar_PCA}. However, comparable considerations jointly spanning both risk and utility dimensions remain limited.
A comprehensive assessment therefore requires methods that jointly account for this multivariate structure, moving beyond simple pairwise comparisons to enable simultaneous consideration of multiple criteria. By framing risk and utility as a multi-objective optimization problem, we provide tools for more holistic evaluation of data releases -- whether synthetically or traditionally anonymized. 
 
In this article, we extend the classical risk-utility map to a multivariate setting through systematic comparison of six visualization approaches: heatmaps, dot plots, composite scatterplots, parallel coordinate plots, radial profile charts, and PCA-based biplots. We introduce three methodological innovations using principal component analysis: (1) blockwise PCA that extracts principal components separately for risk and utility measure blocks, (2) alignment analysis that validates dimensionality reduction by correlating composite measures with principal components, and (3) PCA-based biplots that simultaneously visualize method performances and measure relationships. To our knowledge, this represents the first systematic application of PCA to risk-utility visualization in statistical disclosure control. We further demonstrate systematic Pareto-optimal approach identification across all visualization approaches and evaluate each approach across twelve evaluation criteria. The design of these tools follows Tufte's data-ink principle \parencite{Tufte_1983_Visual} and key guidelines for 
visual data communication \parencite{Franconeri_2021_Science}.

The remainder of the paper is organized as follows. Section~\ref{sec:methodology} introduces Pareto optimality and discusses key considerations related to the scaling of risk and utility measures. Section~\ref{sec:viz_tools} describes the example data used for illustration, presents the visualization methods, and discusses their applications; a systematic comparison of the methods is provided in Section~\ref{sec:method_overview}. Section~\ref{sec:discussion} explores practical implications and outlines directions for future research. Finally, Section~\ref{sec:conclusion} summarizes the main contributions of the paper and offers key recommendations for practitioners.
The visualization methods are presented in order of increasing analytical depth, from tabular overviews to dimensionality-reduction-based approaches.

\section{Methodological Framework}\label{sec:methodology}

\subsection{Pareto-Optimal/Efficient Trade-offs}\label{sec:pareto}
The multivariate evaluation of disclosure risk and data utility naturally constitutes a multi-objective optimization problem, since improvements in one dimension often come at the expense of another. In such settings, the objectives are said to be at least partly conflicting \parencite[p.~5]{Miettinen_1998_Nonlinear}, and a solution (or observation) is called non-dominated, or Pareto-optimal/Pareto-efficient, if none of the objectives can be improved without degrading at least one of the others. Intuitively, this means that a Pareto-optimal solution represents an efficient trade-off: it cannot be improved in one aspect (e.g., utility) without performing worse in another (e.g., risk). Any method that is not Pareto-optimal is strictly inferior to at least one alternative \parencite[p.~547]{MasColell_1995_Chapter}, a concept originating with Vilfredo Pareto's \textit{Cours d'économie politique} (1896--1897).

Formally, let $\mathbf{u}_i \in \mathbb{R}^p$ denote the vector of $p$ utility measures and $\mathbf{r}_i \in \mathbb{R}^q$ the vector of $q$ risk measures for an anonymization approach $i$. We say that anonymization approach $i$ dominates anonymization approach $j$ if 
\[
u_{ik} \geq u_{jk} \quad \text{for all } k=1,\dots,p, 
\quad \text{and} \quad
r_{i\ell} \leq r_{j\ell} \quad \text{for all } \ell=1,\dots,q,
\]
with at least one strict inequality \parencite[p.~11]{Miettinen_1998_Nonlinear} \parencite[Def.~5, p.~588]{Emmerich_2018_Tutorial}. An anonymization approach $i$ is Pareto-optimal if there exists no other anonymization approach $j$ that dominates it. This corresponds to strong Pareto optimality, since at least one inequality must be strict; the weaker notion, which also allows ties, is not considered here \parencite[p.~19]{Miettinen_1998_Nonlinear}.

The collection of all non-dominated solutions forms the Pareto frontier, which makes the trade-off between risk and utility explicit by identifying efficient configurations and distinguishing them from dominated alternatives. Without additional preference information, there may exist an infinite set of Pareto-optimal solutions, all formally equally valid.

Additional decision criteria can also be applied. One option is to impose a risk-tolerance threshold and select the Pareto-optimal solution with the highest utility subject to this bound. Another intuitive “bang-for-buck” heuristic, applicable when the frontier is smooth and concave toward the origin, is to choose the knee (or elbow) point -- where marginal increases in risk begin to yield only diminishing gains in utility, corresponding to the location of greatest curvature along the curve \parencite[p.~275]{Thorndike_1953_Who}.

In practice, Pareto-optimal observations can be highlighted in visualizations using different colors or shapes \parencite{Nagar_2023_Visualization}, but the frontier itself can only be directly visualized in two dimensions. Pareto-optimality can be evaluated either on the full vector of raw measures or on aggregated composite scores.
The former is theoretically cleaner -- strictly monotone transformations preserve dominance metric-by-metric -- but inconsistent with two-dimensional visualization and prone to degeneracy: with many measures, it becomes increasingly difficult for any approach to dominate another on all dimensions simultaneously, causing most approaches to appear non-dominated. The latter reduces the problem to two dimensions at the cost of making the identified set sensitive to aggregation and scaling choices.

\subsection{Orientation and Scaling of Risk and Utility Measures}\label{sec:scaling}

Since risk and utility measures capture fundamentally different aspects of data properties (e.g., disclosure probabilities and distance distributions), they are naturally defined on different scales, units, and directions. Whether higher values indicate better or worse outcomes depends on the measure's definition: for instance, disclosure risk may quantify the attacker's success (higher = worse) or the defender's success (higher = better), and distributional utility may be expressed as distance (lower = better) or as similarity (higher = better). Moreover, measures differ in range and units -- some are bounded probabilities on $[0,1]$, others are unbounded distances -- making direct comparison across measures impossible without rescaling. For joint multivariate analysis and visualization, both a consistent orientation and a transformation to a common scale are therefore necessary.

When measures within a category have mixed orientations (e.g., some utility measures are distance-based where lower = better, while others are similarity-based where higher = better), we recommend reorienting each individually before scaling. Whether risk and utility should share a common direction depends on the intended visualization: 
separate orientations are natural for dual-panel displays such as dot plots (Section~\ref{sec:theory_dotplot}), split parallel coordinate plots (PCPs; Section~\ref{sec:theory_pcp}) and heatmaps (Section~\ref{sec:theory_heatmaps}) with direction-encoded color scales, while a unified direction is recommended for combined radar charts (Section~\ref{sec:theory_radial}), single-panel PCPs, and threshold-based scaling approaches where a single threshold should carry the same interpretation across all axes. When all measures within each category already share the same direction, a single reversal, $x' = 1 - x$ after min--max scaling, or $x' = -x$ after z-score standardization, can reorient an entire category (i.e., utility or risk) for intuitive visual interpretation.

Multivariate visualizations also require comparable axis ranges. Radar charts need bounded scales to produce meaningful polygon shapes, parallel coordinate plots benefit from comparable ranges for visual readability, and PCA biplots require standardization to prevent variables with larger ranges from dominating the principal components \parencite{Greenacre_2010_Biplots}. The choice of scaling method depends on the analytical objective, and conventional normalization approaches do not always yield high-quality visual or structural representations \parencite{Dierkes_2025_Scalingevariant}. Min--max scaling to $[0,1]$ is particularly suitable for radar charts and parallel coordinate plots, as 0 and 1 correspond to the observed worst and best performance across synthetic data generators (SDGs), directly conveying the attainable performance range. z-score standardization is generally preferred for PCA biplots, as it equalizes variances across variables. However, z-score values express performance relative to the mean in standard deviation units, which is less intuitive for visualizing performance trade-offs than the bounded $[0,1]$ range of min--max scaling.

Both of these transformations necessarily abstract from the original metric units, which implies that absolute interpretability of thresholds -- such as maximum acceptable disclosure risk levels -- cannot be directly represented in the scaled space. 
Alternative approaches that explicitly incorporate such acceptability criteria include desirability functions \parencite{Derringer_1980_Simultaneous}, which map raw values to a [0,1] scale using threshold-anchored transformations, and sigmoid functions, which provide smooth bounded mappings centered around a target value. While some measures have well-established baselines, thresholds for many synthetic data quality measures are context-dependent and must be defined for the specific use case and data environment \parencite{Drechsler_2022_Challenges}. Table~\ref{tab:scaling} compares these approaches across four desirable scaling properties -- boundedness, a universal threshold, full differentiation, and fixed variance -- illustrating that no single approach satisfies all criteria simultaneously.
\begin{table}[ht]
\centering
\small
\begin{threeparttable}
\caption{Desirable scaling properties}
\label{tab:scaling}
\begin{tabular}{lccccc}
\toprule
\makecell{Scaling\\approach} & Formula & \makecell{Bounded\\$[0,1]$} &
\makecell{Universal\\threshold\tnote{$\dagger$}} &
\makecell{Full\\differentiation} &
\makecell{Fixed\\variance} \\
\midrule
Min--max & $\frac{x - x_{\min}}{x_{\max} - x_{\min}}$ &
\checkmark & $\times$ & \checkmark & $\times$ \\[6pt]
z-score & $\frac{x - \bar{x}}{s}$ &
$\times$ & $\times$ & \checkmark & \checkmark \\[6pt]
Threshold-proportional & $\frac{x}{\tau}$ &
$\times$ & \checkmark & \checkmark & $\times$ \\[6pt]
Sigmoid & $\frac{1}{1 + e^{-k(x - \tau)}}$ &
\checkmark & \checkmark & $\sim$\tnote{*} & $\times$ \\[6pt]
Desirability & $d_i(x_i)$ &
\checkmark & \checkmark & $\times$\tnote{**} & $\times$ \\
\bottomrule
\end{tabular}
\begin{tablenotes}\tiny
\item $\tau$: acceptability threshold; $k$: steepness parameter;
$d_i$: individual desirability function.
\item[*] Soft approximation; loses differentiation in the tails.
\item[**] Differentiates within the acceptable range;
assigns zero outside it.
\item[$\dagger$] Requires consistent orientation of all measures
so that $\tau$ carries a uniform interpretation across all axes.
\end{tablenotes}
\end{threeparttable}
\end{table}

When Pareto dominance is evaluated directly on the full vector of risk and utility measures, strictly monotone transformations of individual metrics do not alter the set of non-dominated solutions, since dominance is defined through metric-by-metric comparisons and strictly monotone transformations preserve the ordering within each metric. However, the orientation of each measure must be consistent with the optimization direction. When dominance is instead assessed on aggregated composite scores (e.g., mean risk and mean utility), scaling implicitly determines the relative contribution of each measure to the composite, and consequently the identified Pareto set may differ depending on the chosen scaling transformation (see Appendix~\ref{app:pareto} for a numerical illustration).

\section{Visualization and Evaluation Tools}\label{sec:viz_tools}
This section develops a visual approach to the multi-objective optimization problem underlying the disclosure risk-utility trade-off. We introduce methods well suited to evaluating and visualizing anonymized and synthesized data across multiple risk and utility dimensions. As argued in Section~\ref{sec:intro}, the risk-utility problem is fundamentally multivariate. We therefore introduce multivariate visualization strategies: different PCA biplots, R-U maps, origami plots and related multivariate views with Pareto-optimal solutions highlighted.

To demonstrate the proposed visualization methods, data from the European Union Statistics on Income and Living Conditions (EU-SILC) are utilized. EU-SILC is a flagship data source in official statistics, widely used to monitor poverty, social inclusion, and related policy targets in Europe. We use the Austrian EU-SILC public-use file from 2013. The data has a hierarchical structure, with individuals nested within households.  A concise description of key variables and further details is available in the manual of the \texttt{R} package \texttt{simPop} \parencite{Templ_2017_Simulation}; a comprehensive variable catalogue is provided in \textcite{silc} and online at \url{https://www.gesis.org/en/missy/materials/EU-SILC/documents/codebooks} (accessed 2025-06-25).

We synthesize data using a methodologically diverse set of commonly used synthetic data generators (SDGs), each applied 10 times to account for stochastic variation. The SDGs and their software packages are described in Table \ref{tab:SDGs} in Appendix~\ref{app:SDGs}; SDG-specific parameter settings used for the EU-SILC synthesis are provided in Table \ref{tab:sdg_settings_eusilc} of the same appendix.

Over the past 30 years, the field of synthetic data has grown substantially, giving rise to a large number of evaluation metrics \parencite{Kaabachi_2025_Scoping}. However, no clear consensus has emerged on a standard set of measures \parencite{Drechsler_2024_30}. We therefore select a diverse range of metrics spanning different categories of both utility and risk. For utility, we draw on the taxonomy of \textcite{Drechsler_2024_30} and choose among global utility measures (e.g. pMSE), outcome-specific utility measures (e.g. Confidence Interval Proximity), and fit-for-purpose measures (e.g. invalid Households with only children). For risk, we draw on the recently published consensus study on privacy metrics for synthetic data \parencite{Pilgram_2025_Consensus}, covering three disclosure categories: identity disclosure (RepU, DCR), attribute disclosure (TCAP, RAPID), and membership inference (MIA). These measures and their descriptions are provided in Table \ref{tab:measure_meaning} (for further details, see Appendix \ref{app:SDGs}, Table \ref{tab:measure_specification}).The MIA implementation is based on the \texttt{SynthEval} framework \parencite{Lautrup_2025_Syntheval}.

We emphasize that the chosen SDGs and evaluation metrics are intended solely to support the proposed visualizations and should not be interpreted as a systematic comparison or benchmarking of different SDGs. The selection of both generators and metrics was made subjectively and is not central to the contribution of this paper, which focuses on proposing and evaluating visualizations for the multivariate nature of the risk-utility trade-off. Accordingly, the SDGs were not extensively tuned to the specific dataset, and the results should not be interpreted as performance comparisons between SDGs.

\begin{table}[htbp]
\centering
\caption{Description of the risk and utility measures.}
\label{tab:measure_meaning}
\normalsize
\begin{threeparttable}
\begin{adjustbox}{width=\textwidth}
\begin{tabular}{l p{8cm} p{6.5cm} p{9.5cm}}
\toprule
\textbf{Type} & \textbf{Measure \& Abbreviation} & \textbf{Author / Used in} & \textbf{Description} \\
\midrule
Risk    & Replicated Uniques (\texttt{RepU})                          & \citep[e.g.,][]{Raab_2025_Practical, Raab_2025_Confidentiality}                                    & Percentage of replicated sample uniques in synthetic dataset. \\
Risk    & Distance to Closest Record (\texttt{DCR})                   & \citep[e.g.,][]{Yao_2025_DCR}                                           & Ratio of mean nearest-neighbour distances from synthetic records to training data versus holdout.\\
Risk    & Membership Inference Attack (\texttt{MIA})                  & \citep[e.g.,][]{Lautrup_2025_Syntheval, ElEmam_2022_Validating, Houssiau_2022_TAPAS, Shokri_2017_Membership}         & Probability that an attacker can correctly infer whether a record was part of the training data. \\
Risk    & Targeted Correct Attribution Probability (\texttt{TCAP})    & \citep{Taub_2019_Creating}                                     & Probability of correct attribute inference. \\
Risk    & Risk of Attribute Prediction-Induced Disclosure (\texttt{RAPID}) & \citep{Templ_2026_RAPID}                                  & Expected share of predicted attribute disclosure. \\
\midrule
Utility & Confidence Interval Proximity (\texttt{CIProx})          &   \citep[e.g.,][]{Karr_2006_Frameworka, Drechsler_2009_Disclosure}                                                              & Deviation of confidence intervals for a sensitive variable. \\
Utility & Propensity Mean Squared Error (\texttt{pMSE})               & \citep[e.g.,][]{snoke_general_2017, Raab_2021_Assessing}                                     & Predictive score from distinguishing real vs.\ synthetic data. \\
Utility & Wasserstein Distance (\texttt{Wasserstein})                 & \citep{Vasershteyn_1969_Markovskie}                            & Wasserstein distance between numeric distributions. \\
Utility & Households with only children (\texttt{NoAdultHH}) & \citep[e.g.,][]{thees24} & Households in which all members are under 18 years of age. Since such households are demographically impossible in the EU-SILC data structure, the count should be zero in valid synthetic data. A non-zero value indicates a logical consistency violation.\\
Utility & Correlation Matrices Differences (\texttt{CMD})             &         \citep[e.g.,][]{Miletic_2025_Utilitybased}                                            & Mean absolute difference between corresponding real and synthetic correlation coefficients. \\
\bottomrule
\multicolumn{4}{l}{\footnotesize\textit{Note:} CIProx adapts the confidence interval overlap measure, applying it to variable-level means rather than regression estimands.}
\end{tabular}
\end{adjustbox}
\end{threeparttable}
\end{table}

Regarding orientation, all utility measures are aligned such that higher values indicate better performance. For risk measures, lower values indicate lower disclosure risk; an exception is the Distance to Closest Record (DCR), which by construction yields higher values for lower risk and was therefore inverted to ensure a consistent orientation across all risk measures.

This orientation is maintained throughout the analysis. For certain visualizations that require a unified interpretation of radial extent (e.g., origami plots), risk measures are temporarily inverted so that larger values consistently correspond to better performance across all axes; this transformation is applied solely for visual interpretability and does not affect any quantitative analysis.

For scaling, different transformations are used depending on the analytical objective. Min–max scaling to the interval [0,1] is applied for composite R–U maps and bounded multivariate visualizations such as heatmaps, dot plots, and radial charts, where interpretability of the observed performance range is desirable. Scaling is performed per metric across all SDGs. 

Pareto-optimality is assessed on the min–max scaled composite scores -- mean utility and mean risk. Although min–max scaling is sensitive to extreme values, the set of SDGs in our setting is fixed, and extreme observations -- including the original dataset as a reference point -- represent meaningful benchmark anchors rather than nuisance outliers. For PCA-based visualizations, z-score standardization is applied to ensure comparability across measures.

To verify this choice, we compared PCA biplots under min--max scaling, z-score standardization, and the scaling optimization algorithm proposed by \textcite{Dierkes_2025_Scalingevariant}, which uses Nelder--Mead optimization to determine per-dimension scaling factors that maximize a visual quality criterion for two-dimensional projections. Min--max and z-score scaling produced broadly similar projections with comparable cluster structure and synthesizer separation, confirming that the visualization is robust to the choice between these two common scaling approaches for the present data. The Dierkes optimization, applied using the \textit{Data Space Ratio} as the unsupervised quality criterion, collapsed the projection to a near-one-dimensional solution and was therefore discarded. We proceed with min–max scaling for bounded visualizations and z-score standardization for PCA, as described above.
All risk and utility measure calculations, data manipulation, and visualization were performed in \texttt{R} \parencite{RCoreTeam_2025_Language}\footnote{We mainly used the \texttt{tidyverse} ecosystem \parencite{Wickham_2019_Welcome}, specifically \texttt{ggplot2} \parencite{Wickham_2016_Ggplot2} for visualization. Additional packages are cited where used.}.

\subsection{Heatmaps}\label{sec:theory_heatmaps}

As an initial approach to visualizing multiple risk and utility metrics simultaneously, we present heatmaps. Heatmaps represent the values of a data matrix by encoding them as color intensities \parencite{Wilkinson_2009_History_heatmap}. In the context of statistical disclosure control, they provide a compact overview of how multiple anonymization strategies perform across a range of risk and utility measures.  

\begin{figure}[H]
    \centering
    \includegraphics[width=1\linewidth]{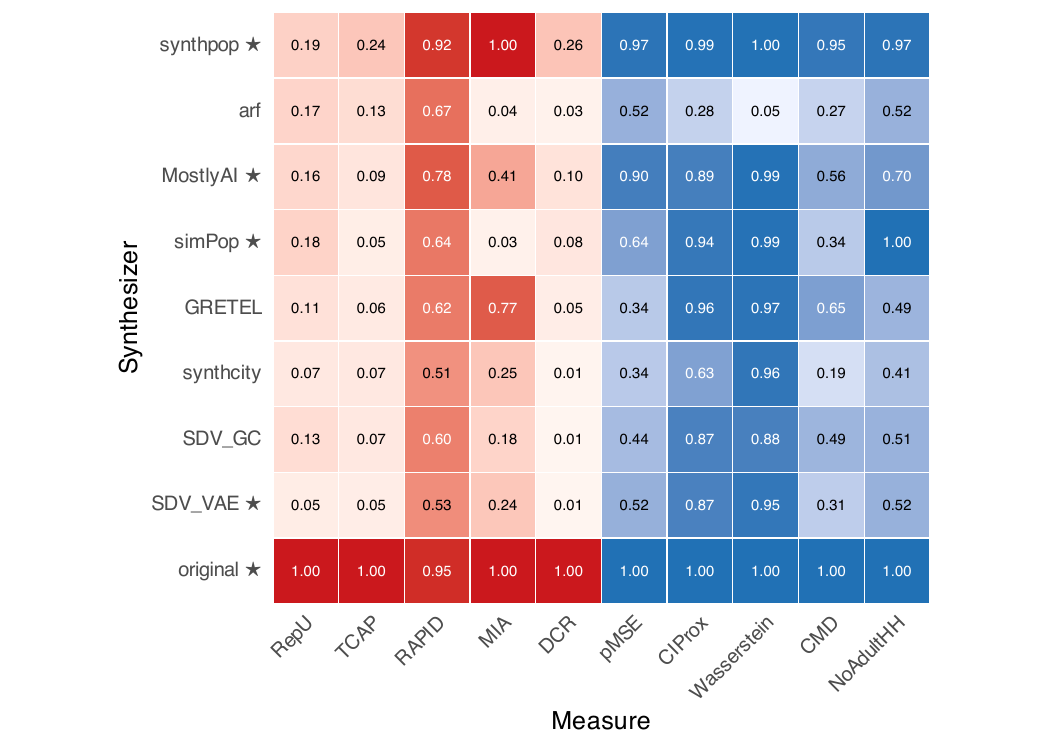}
    \caption{Performance of synthetic data generators across utility (bluish) and risk measures (reddish), with values min--max normalized to [0,1] (rows = SDGs, columns = measures). Asterisks indicate composite Pareto-optimal SDGs.}
    \label{fig:heatmap}
\end{figure}

Two main variants can be distinguished. In the first form, used in Figure \ref{fig:heatmap}, the heatmap displays anonymization approaches as rows and evaluation measures as columns. Each tile then shows the performance of one method on one risk or utility metric (mean of 10 synthetic datasets per SDG), with darker or lighter colors indicating higher or lower values, depending on the scale. This layout facilitates visual comparison both across metrics (horizontally) and across anonymization strategies (vertically). When multiple datasets are evaluated, summarizing like in Figure \ref{fig:heatmap} or faceting and small multiples can be used to assign one panel per dataset, supporting cross-dataset comparisons. Hierarchical clustering may be applied to reorder rows or columns based on similarity, thereby highlighting patterns or grouping structures. 

In an alternative representation, each heatmap corresponds to a single anonymization method, with utility measures along one axis and risk measures along the other. Here, the tile color reflects the performance on a particular utility–risk pair. This variant highlights joint behavior across metrics and can be useful for examining interaction structures or identifying imbalanced profiles (e.g., methods that perform well in utility but poorly in specific risk dimensions). However, comparisons between methods are then made across different plots, rather than within a single unified display. 

In both formats, enhancements such as numeric value labels, normalization per measure, or visual markers for Pareto-optimal methods (e.g., asterisks or outlines) can improve interpretability. Despite their strengths, heatmaps represent only univariate values per tile, so inter-metric correlations and higher-order patterns remain hidden. Colour encoding may distort perception, especially when a few large values compress the visual range of the smaller ones. 

Figure \ref{fig:heatmap} presents a clustered heatmap of min--max normalized measures to reveal cross-method patterns at a glance. To indicate Pareto-optimal SDGs we applied a standard two-objective dominance rule (minimizing Risk, maximizing Utility) explained in Section \ref{sec:pareto}. That is, for SDGs $i$ and $j$ (i.e., their mean risk and utility scores over the 10 synthetic datasets), we say that $j$ dominates $i$ if $\mathrm{Risk}_j \le \mathrm{Risk}_i \quad \text{and} \quad \mathrm{Utility}_j \ge \mathrm{Utility}_i$,
with at least one strict inequality. SDGs that are not dominated by any other form the Pareto set and are indicated with a star in Figure~\ref{fig:heatmap}. For the row-wise ordering of the SDGs, hierarchical clustering was used. The heatmap shows that \texttt{MostlyAI}, \texttt{synthpop}, \texttt{SDV\_VAE} and \texttt{simPop} are the only SDGs that are on mean basis Pareto-optimal.

\subsection{Dot Plots and Distributional Extensions}\label{sec:theory_dotplot}

Dot plots provide a position-based alternative to heatmaps by encoding values as points on a common scale. Faceting separates risk and utility metrics, and 
Pareto-optimal methods can be highlighted through shape or color coding. Position-based encodings allow more accurate magnitude comparisons than color-based representations, as perceptual judgments of length and alignment 
are more reliable than those of hue or intensity \parencite{Cleveland_1984_Graphical}, facilitating the detection of outliers, skewed profiles, and uneven metric contributions.

\begin{figure}[H]
    \centering
    \includegraphics[width=1\linewidth, clip]{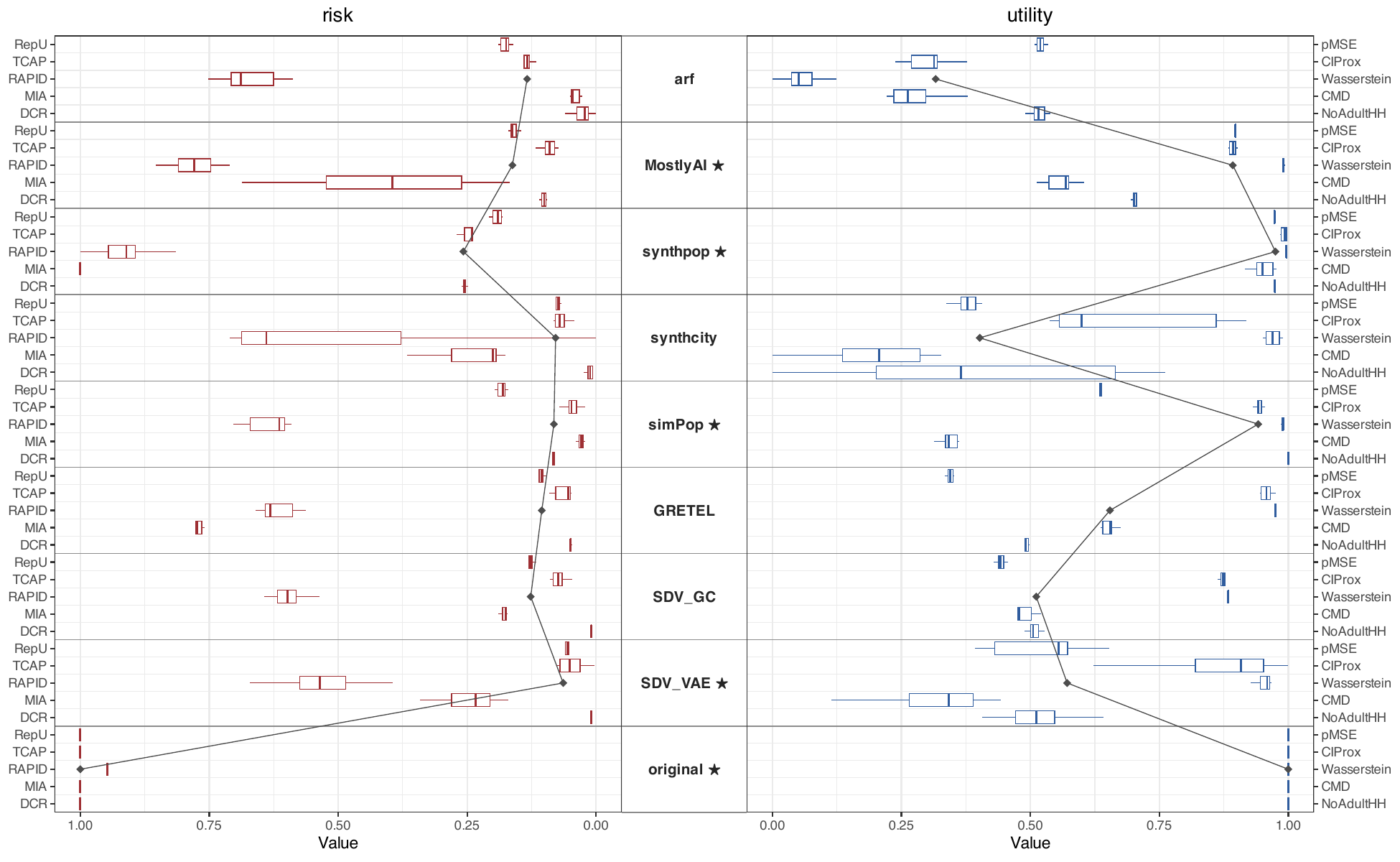}
\caption{Distributional dot plot with box plots of risk (left, red) and utility (right, blue) measures across synthesizers. Boxplots display the distribution of individual measure values across replications, while the grey diamonds and connecting lines indicate the per-synthesizer median across measures.}
    \label{fig:dotplot}
\end{figure}

When each measure is observed repeatedly (e.g., across multiple synthetic datasets or repeated runs), dot plots can be extended with boxplots to summarize 
within-method dispersion via the median and interquartile range, revealing instability or occasional extreme outcomes. When only a single value per method 
is available, the visualization reduces to a classical dot plot. Figure~\ref{fig:dotplot} shows this distributional extension.

\subsection{Composite Risk/Utility Scatterplots}\label{sec:theory_ru_map}

\begin{figure}[H]
    \centering
    \makebox[\textwidth][c]{\includegraphics[width=1.4\linewidth]{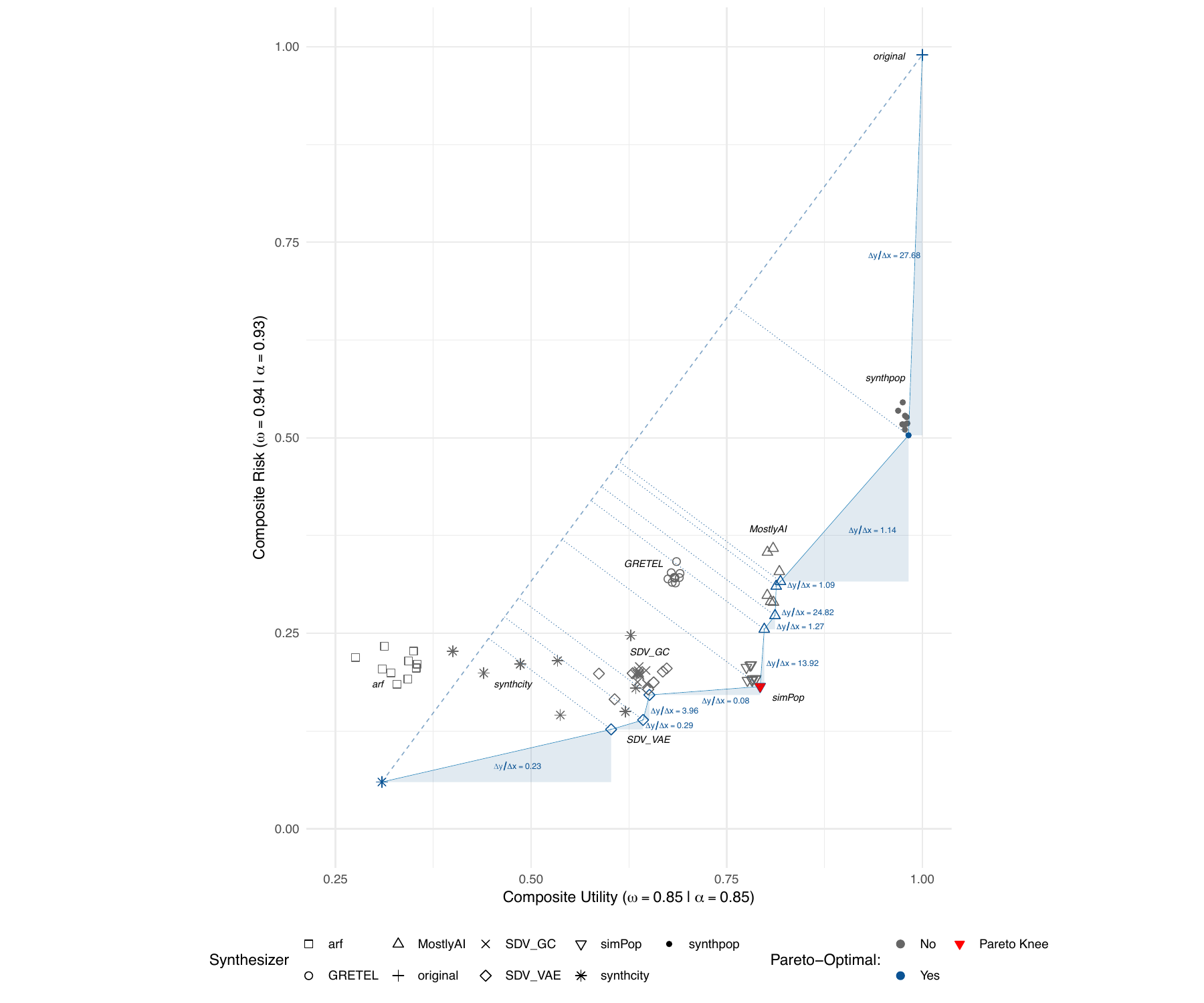}}
    \caption{Composite risk vs. utility for SDGs. Pareto-optimal methods are in blue; reliability of composites is indicated by  $\alpha$ and $\omega$. 
    }
    \label{fig:composite_RU_extended}
\end{figure}

Composite scatterplots summarize the performance of different anonymization approaches by reducing multiple risk and utility measures into two composite scores \parencite[see, e.g.,][]{Little_2022_Comparing}. The horizontal axis shows composite utility, e.g., an average of all utility measures, while the vertical axis shows composite risk, similarly an average of all risk measures. Each point in the scatterplot corresponds to an anonymization approach. The goal of the plot is to highlight trade-offs between disclosure risk and data utility, making it possible to identify anonymization strategies that strike a good balance. The quadrant structure is the same as in a classical R-U map and provides an intuitive interpretation (see Figure \ref{fig:composite_RU_extended}):

\begin{itemize}
         \item Bottom-right: desirable region -- high utility with low risk (best trade-off).
         \item Top-right: high utility but also high risk, posing disclosure concerns.
         \item Bottom-left: low utility but also low risk, often of little practical value.
         \item Top-left: worst case -- low utility combined with high risk.
\end{itemize}

    The strengths of this visualization are its simplicity and interpretability: it reduces an $n$-dimensional evaluation to two axes, enabling a quick diagnostic view and straightforward comparisons across anonymization strategies. At the same time, several limitations need to be acknowledged. Collapsing all measures into averages inevitably leads to information loss, hides variability across metrics, and implicitly assumes equal weighting of measures. The approach also ignores metric correlations and it does not display uncertainty such as standard errors or confidence intervals.  Finally, results may depend on whether measures have been standardized prior to aggregation.
    
    To enhance interpretability, the composite R–U map can be augmented with several features. First, highlighting the composite Pareto-optimal set separates genuinely optimal anonymization solutions from dominated ones, transforming the plot from a summary into a decision aid. Second, local trade-off annotations reveal the marginal "price" of utility: for each anonymization approach, showing the incremental move to the next better one (e.g., labels with $\Delta$ Utility and $\Delta$ Risk) indicates how much additional disclosure risk is incurred per unit of utility gained. Third, error bars showing the standard deviation of risk and utility along their respective axes indicate dispersion across the underlying measures. Fourth, a horizontal risk-tolerance threshold line (e.g., at a maximum 
    acceptable composite risk level $\bar{r}_{\max}$) can in principle be overlaid to partition the plot into an acceptable region (below the line) and an unacceptable region (above). However, we do not recommend this in practice: individual risk measures rarely have universally agreed thresholds, and aggregating them into a composite further obscures interpretation.

While composite scores are commonly used to summarize performance across multiple risk or utility metrics, their interpretability hinges on the degree to which the constituent measures reflect a coherent underlying construct. To assess this, internal consistency metrics such as Cronbach’s $\alpha$ or McDonald’s $\omega$ can be used as heuristic diagnostics \parencite{Cronbach_1951_Coefficient,McDonald_1999_Omega,Zinbarg_2005_Cronbachs,Hayes_2020_Use}. A value of $\alpha \geq 0.70$ is often cited as indicative of acceptable consistency among items within a block {\parencite[p.~230]{Taber_2018_Use}; \parencite[p.~245]{Nunnally_1994_Psychometric}}. A high value then means that the composite score is a reliable summary of the underlying set of risk or utility measures, reflecting shared variation rather than averaging over unrelated or inconsistent metrics. Note that the mentioned threshold is a rule-of-thumb and context-dependent, and $\alpha$ assumes tau-equivalence -- i.e., equal contributions of all indicators to the latent construct -- which may not hold in practice. McDonald’s $\omega$ is more general, as it allows heterogeneous loadings across items. In the context of multivariate risk–utility evaluation, internal consistency estimates serve only as rough checks for whether averaging across metrics (to form composite scores) is justifiable. 

Figure~\ref{fig:composite_RU_extended} shows an exemplary composite R–U map. For 10 iterations per SDG we compute composite scores by averaging the normalized constituent measures separately for risk (lower is better) and utility (higher is better). The resulting scatterplot places each SDG in the R-U plane. We report Cronbach’s~$\alpha$ and McDonald’s~$\omega$ for the sets of metrics entering each composite; both indicate that the composites capture a coherent underlying construct. Pareto-optimal SDGs are shown in blue and connected to form the empirical Pareto front. We highlight \texttt{simPop} as the knee point (Section \ref{sec:pareto}), defined as the point on the front with the largest perpendicular distance from the straight line joining the two extreme Pareto points, which approximates the location of maximum curvature \parencite[p.~275]{Thorndike_1953_Who}. Further we display the slope values $\Delta R/\Delta U$, which represent the trade-off from one SDG to another on the Pareto front.  
 
\begin{figure}[H]
    \centering
    \includegraphics[width=1\linewidth]{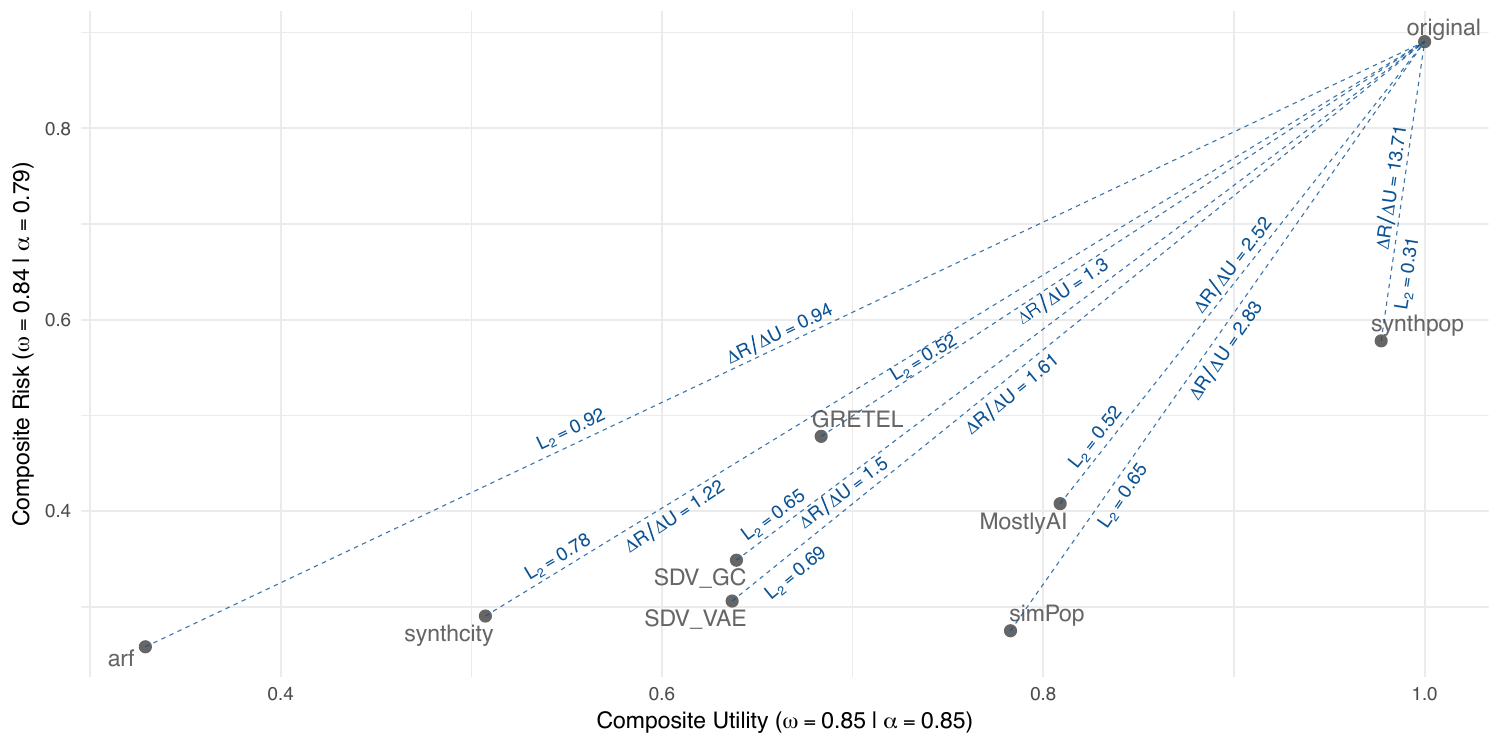}
    \caption{Rays from each SDG mean across 10 iterations to the original $(U_0,R_0)$ with labels $\Delta R/\Delta U$ and Euclidean distance $L_2$. 
    }

    \label{fig:slopes_RU}
\end{figure}

In Figure~\ref{fig:slopes_RU} we draw, for each mean SDG point $(U_i,R_i)$ across 10 iterations, the ray to the original $(U_0,R_0)$ and report the Euclidean $L_2$ distance and the slope
\[
\text{slope}_i^{(\mathrm{orig})}=\frac{\Delta R}{\Delta U}
=\frac{R_i - R_0}{\,U_i - U_0\,},
\]
interpreted (for $\Delta U>0$) as the incremental risk paid per unit of utility gained relative to the original. A lower $\text{slope}_i^{(\mathrm{orig})}$ means a cheaper marginal trade-off from the original, but it is not an overall ranking: global desirability depends on the joint $(U_i,R_i)$ and the full metric set. A similar picture is revealed when we look at simple Euclidean distance ($L_2$) in the R-U plane. Without explicit scaling/weighting, these can be misleading; for example, \texttt{SDV\_GC} and \texttt{simPop} can be equally distant from the original even though the latter Pareto-dominates the former in terms of the mean. We therefore use Pareto dominance to identify “best”, and use slopes only to summarize marginal cost.

\subsection{Parallel Coordinate Plot (PCP)}\label{sec:theory_pcp}

Parallel coordinate plots \parencite[PCP;][]{Inselberg_1985_Plane} are a classical technique for visualizing multivariate data and are frequently used in many-objective optimization \parencite{Nagar_2023_Visualization}. They are therefore well suited to the joint analysis of disclosure risk and utility measures.

\begin{figure}[H]
    \centering
    \includegraphics[width=1\linewidth]{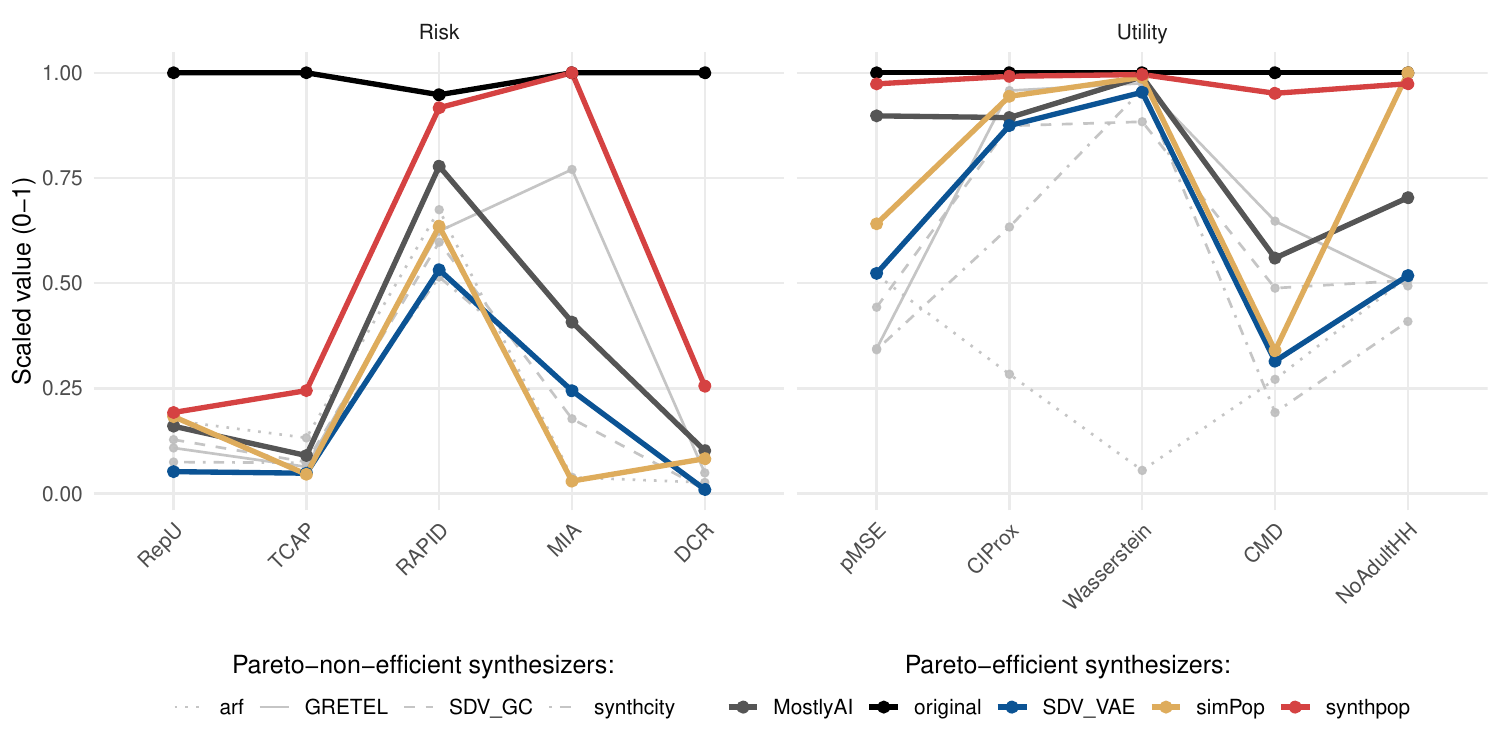}
    \caption{Parallel coordinates of scaled risk (left) and utility (right) measures for SDGs. Pareto-optimal SDGs (colored) are contrasted with non-optimal ones (gray), with the original dataset shown as a benchmark.} 
    \label{fig:pcp}
\end{figure}

In a PCP, each SDG is represented as a polyline intersecting a series of parallel vertical axes, where each axis corresponds to a min–max normalized evaluation metric (cf. Figure~\ref{fig:pcp}). The vertical position on each axis reflects the scaled value of the respective measure. By connecting these positions, the plot encodes the full multivariate performance profile of each method without reducing it to a single aggregate score.

Figure~\ref{fig:pcp} displays the risk and utility measures (scaled to $[0,1]$) across the SDGs. In contrast to the distributional dot plot, which visualizes dispersion across iterations, the PCP shows the mean value over ten synthetic data iterations, resulting in a single aggregated performance profile per SDG. To account for their conceptual and directional differences, risk and utility are shown again in separate facets. Pareto-optimal SDGs are highlighted in color, while non-efficient methods are rendered in neutral grey for contextual comparison. 

PCPs are particularly useful for revealing overall performance profiles: relatively flat polylines indicate balanced behavior across metrics, whereas pronounced peaks or troughs suggest specialization or weaknesses in specific dimensions. In contrast to scalar summary indicators, this representation preserves the multivariate structure of the evaluation.

When multiple datasets or repeated runs are analyzed, PCPs can quickly become visually dense if all profiles are overlaid. Two common strategies exist to address this: either aggregating results (e.g., by displaying mean profiles, as done in Figure \ref{fig:pcp}) or employing small multiples by faceting. The former reduces clutter through summarization, while the latter preserves full detail across separate but consistently scaled panels.

A known limitation of PCPs is their sensitivity to axis ordering: different sequences of measures along the horizontal axis can alter visual patterns and emphasis. For this reason, we complement PCPs with order-invariant alternatives such as origami plots (cf. Section~\ref{sec:theory_radial}).

\subsection{Radial Profile Plots (Radar / Origami)}\label{sec:theory_radial}

\begin{figure}[H]
    \centering
    \includegraphics[width=\linewidth]{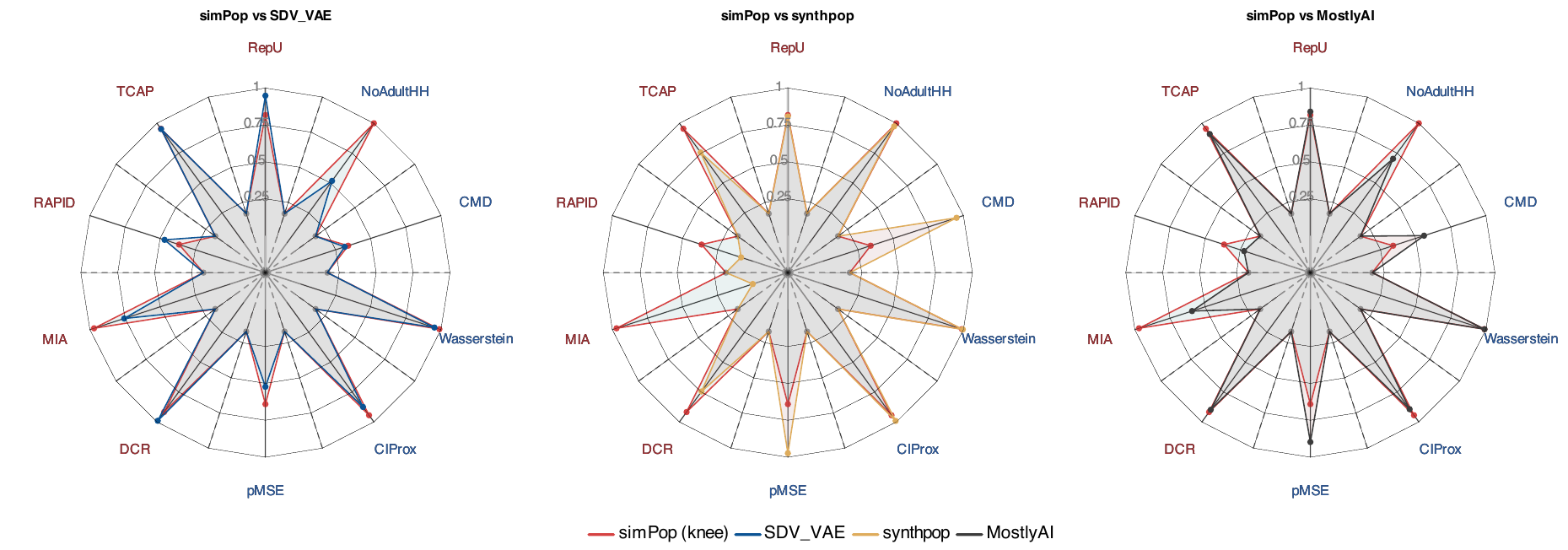}
    \caption{Bivariate origami plots comparing the Pareto-optimal synthesizers against the knee solution (\texttt{simPop}). Each axis represents a normalized performance measure scaled to $[0,1]$, with harmonized orientation such that larger radial distance indicates better performance. Polygon shapes summarize multivariate performance across risk and utility measures, revealing strengths and trade-offs between approaches.
    }
    \label{fig:origami}
\end{figure}

Radial profile plots represent multivariate performance by placing each metric on a separate radial axis and connecting the corresponding values to form a polygonal profile (cf. Figure~\ref{fig:origami}). Similar in spirit to parallel coordinate plots, they provide a compact visual summary of how an anonymization approach performs across multiple risk and utility dimensions simultaneously. Although early forms were developed for cyclic phenomena \parencite[p.~78]{Mayr_1877_Gesetzmaessigkeit}, radial plots are now widely used for general multi-criteria comparison.

Min–max normalized measures are positioned along equally spaced radial axes and connected to form a polygon. The resulting geometry encodes the multivariate performance profile of a method: differences in radial extent across axes reveal strengths and weaknesses on individual metrics, while the overall shape conveys balance or imbalance across dimensions.

A key limitation of classical radar charts is their sensitivity to the ordering of variables around the circle, as different permutations can produce substantially different visual impressions and alter polygonal areas. In classical radar charts, the polygon area between adjacent axes is jointly determined by neighbouring variables, meaning a variable's visual contribution depends not only on its own value but also on those of its neighbours — making comparisons sensitive to arbitrary ordering choices. The Origami plot \parencite{Duan_2023_Origami} addresses this by introducing auxiliary axes arranged such that each variable's contribution to the total polygon area is independent of its position around the circle. The plot is therefore only semi-order-dependent: while the overall polygon shape still varies with variable ordering, each individual spike retains the same geometry regardless of where it is placed around the star. This stabilizes area-based comparisons across methods. The framework additionally allows explicit weighting of measures, enabling emphasis on selected criteria when required.

In line with the previous visualizations, we use aggregated performance profiles per SDG. For visual comparability, risk measures (where lower values indicate better performance) are inverted solely for the origami visualization, ensuring that larger radial extensions consistently correspond to more desirable outcomes across both risk and utility dimensions. This transformation is purely visual and does not affect quantitative results.

Because radial plots quickly become cluttered when many profiles are overlaid, we restrict the visualization to the Pareto-optimal SDGs and present pairwise comparisons against the knee solution (cf.~Section~\ref{sec:pareto}). This preserves readability while enabling direct inspection of structural differences between leading methods.

Figure~\ref{fig:origami} shows these bivariate origami comparisons. The polygonal shapes highlight differences in multivariate composition rather than relying solely on aggregate scores.

Polygon areas can be computed as an additional descriptive summary (cf.~Table~\ref{tab:sdg_area}). To ensure interpretability, all axes are harmonized so that larger values indicate better performance. The comparatively small area of the original dataset illustrates the inherent trade-off structure of the evaluation: perfect utility does not offset maximal disclosure risk once axes are aligned, as the origami area reflects balanced performance across all individual measures.

\begin{table}[H]
\centering
\scriptsize
\caption{Polygon areas of the harmonized origami profiles per synthesizer. Areas are computed from mean-normalized risk and utility measures after aligning axis orientation.}
\vspace{6pt}
\label{tab:sdg_area}
\begin{tabular}{lc}
\toprule
\textbf{SDG} & {\textbf{Area}} \\
\midrule
simPop & 0.79\\
MostlyAI & 0.75\\
SDV\_VAE & 0.73\\
synthpop & 0.73\\
SDV\_GC & 0.72\\
GRETEL & 0.68\\
synthcity & 0.66\\
arf & 0.56\\
original & 0.51\\
\bottomrule
\end{tabular}
\end{table}

\subsection{Multivariate PCA-based R-U Maps}\label{sec:theory_pca}

Multivariate PCA-based R-U maps rely on principal component analysis \parencite[PCA;][]{1901_Pearson, 1933_Hotelling}, which reduces the dimensionality of multivariate data by projecting it into a low-dimensional space while retaining as much variance as possible.  We present two approaches for applying PCA to risk-utility evaluation. The joint PCA approach combines all risk and utility measures in a single analysis and visualizes them in a biplot. The blockwise PCA approach applies PCA separately to risk and utility measure blocks, then plots the resulting principal components in a composite scatterplot.
    
\subsubsection{Joint PCA R-U Map}\label{sec:joint_pca}
    
Biplots, introduced by \textcite{1971_Gabriel}, provide a graphical framework to represent both observations and variables in the same plot. By combining the PCA projection with variable loadings, biplots make it possible to explore patterns, relationships, and group structures in complex datasets. In the context of data anonymization, biplots provide an effective visualization for exploring the trade-off between disclosure risk and data utility while also revealing correlations among multiple evaluation metrics within a single, interpretable representation. Although dimensional reduction methods have previously been used to evaluate the utility of anonymized datasets \parencite[e.g.,][]{2025_Bachot}, we are not aware of prior work applying biplots to assess multiple risk and utility measures simultaneously.

\begin{figure}[H]
    \centering
    \includegraphics[width=\linewidth]{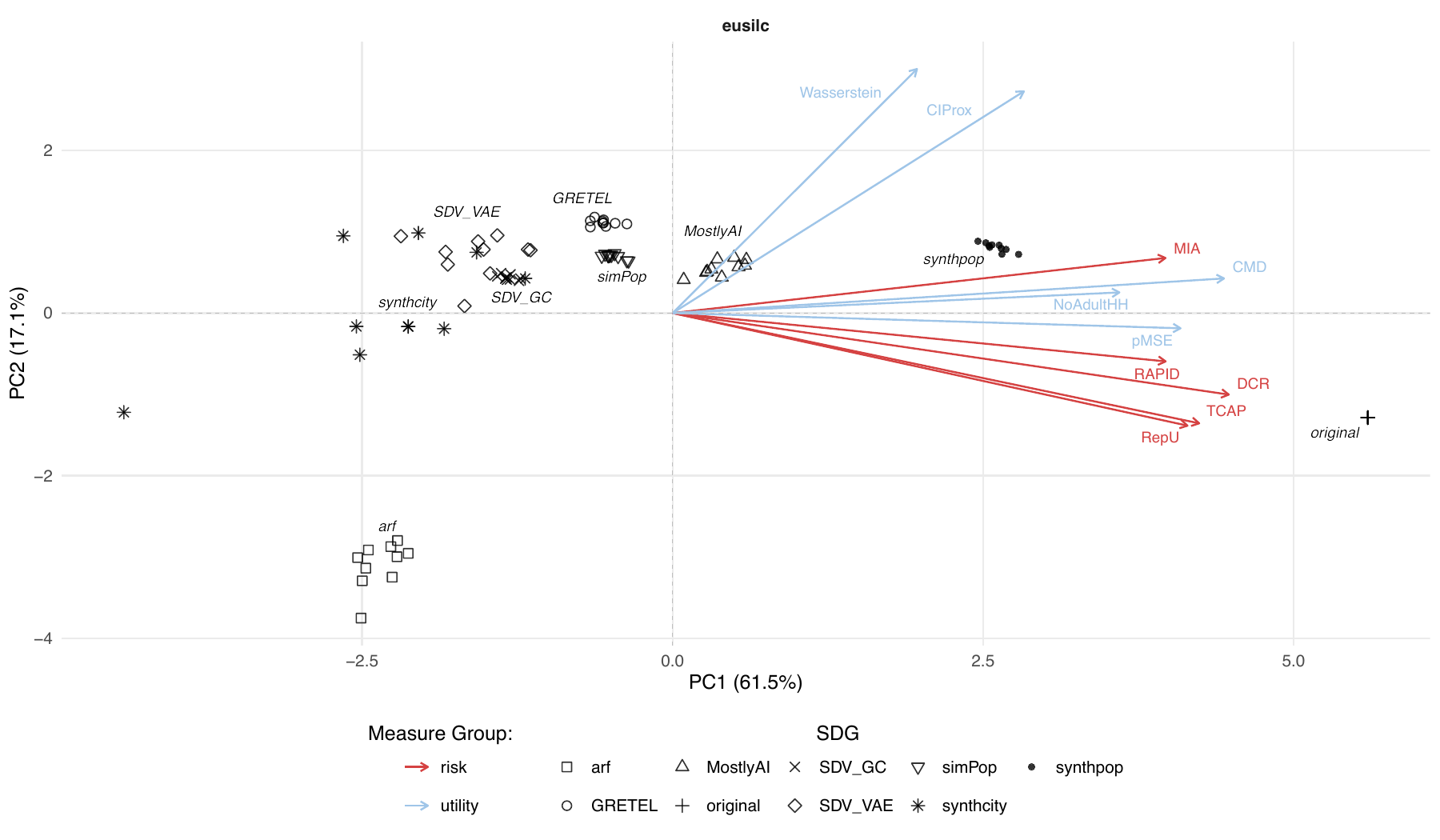}
    \caption{Joint PCA biplot. Points are SDGs (blue = Pareto-optimal); arrows are variable loadings with color indicating measure group (utility vs risk). Arrow length shows a measure’s influence on the PCs; a point’s projection approx. in direction of an arrow indicates association with that measure.}
    \label{fig:joint_pca}
\end{figure}

A biplot based on a joint PCA of risk and utility measures is presented in Figure \ref{fig:joint_pca}. To construct the biplot, PCA is applied to the z-standardized risk and utility measures (cf.\ Section \ref{sec:scaling}), where the first two principal components define the axes on which observations and variable loadings are jointly visualized. In the biplot depicted in Figure \ref{fig:joint_pca}, the first two principal components capture approximately 79\% of the total variation. Each plotted point represents an iteration of an anonymization approach, with shape indicating the SDG. The plot can be further augmented with a point color (e.g., to highlight Pareto-optimal approaches) and confidence ellipses. 
In the biplot, approaches that lie close together have similar risk–utility profiles, while isolated points may indicate outliers (e.g., very low utility and/or very low risk). Loading arrows indicate how the measures correlate, highlighting which risk and utility measures align or contrast. For example, in Figure \ref{fig:joint_pca}, risk measures (red arrows) are strongly aligned with the PC1 axis, where most risk and utility measures are highly positively correlated, reflecting the risk-utility trade-off. In this illustrative example, generators such as \texttt{MostlyAI} and \texttt{synthpop} are associated with high data utility but also with high risk relative to the other generators. The plot additionally reveals groups of SDGs with similar risk-utility profiles as well as outliers like \texttt{arf} with its particularly low utility regarding the measures \textit{Wasserstein} and \textit{CIProx}. When outliers are a concern, a robust PCA variant (e.g., ROBPCA; \cite{Hubert_2005_ROBPCA}) is preferable (see Appendix \ref{app:robPCA} for details). 

Note that the PCA axes are linear combinations of the original measures and may mix risk and utility; interpretation depends on how the measures load onto the principal components, as well as on the standardization applied and the assumption of linearity.

Additionally, principal components are sign-indeterminate: flipping a component and its loadings leaves the general structure of the biplot unchanged. Although one could fix signs in the joint PCA for readability -- e.g., enforce $\operatorname{corr}(PC1,U)\!\ge\!0$ and $\operatorname{corr}(PC2,-R)\!\ge\!0$ and apply the same flips to scores and loadings -- we keep the native orientations returned by the algorithms throughout. To aid interpretation, figures annotate the directions of higher utility ($U$) and lower risk ($R$) and display loading arrows.

\paragraph{Compare anonymization approaches across multiple datasets:} When multiple datasets are anonymized (e.g., to compare the performance of different SDGs across datasets), each point in the biplot represents a dataset--anonymization method pair. The biplot then separates datasets and positions anonymization approaches accordingly. When iterations of the same SDG are plotted individually, as in Figure~\ref{fig:joint_pca}, the resulting clusters visualize within-method variation across runs; the same principle extends to multiple datasets, where per-dataset centroids and ellipses can summarize between-dataset variation for each SDG. This complements joint PCA, alignment, and blockwise PCA (see following Section \ref{sec:blockwise_pca}) for multi-dataset comparisons.

\paragraph{PC1 alignment with risk and utility composites:} To support interpretation, we investigate the loadings in detail and relate the PC1 scores to the composite measures of risk and utility (cf. Table \ref{tab:alignment_eusilc}). The loadings show how risk and utility measures contribute to the principal components, but our broader aim is to assess whether risk and utility as constructs are distinguishable, i.e., whether they occupy separate directions in the principal component space. We therefore correlate the PC scores with the composite risk and utility indicators (each constructed as the mean of their respective measures) to quantify how well each component aligns with these two conceptual dimensions. 

Formally, let $X \in \mathbb{R}^{n \times p}$ denote the data matrix where rows correspond to individual SDG iterations (10 runs per SDG) and columns to risk--utility measures, $x_i \in \mathbb{R}^p$ is the column vector for observation $i$, and $\mu \in \mathbb{R}^p$ is the column vector of variable means across all $n$ observations. With $k = 2$ retained components, the loading matrix is
\[
P_{p,k} = [\,\mathbf{p}_1,\dots,\mathbf{p}_k\,] \in \mathbb{R}^{p\times k},
\]
and the score vector for observation $i$ is
\[
t_i = P_{p,k}^\top (x_i - \mu) \in \mathbb{R}^k,
\]
with PC1 score $t_{i1} = \mathbf{p}_1^\top (x_i - \mu)$. Stacking $t_i^\top$ as rows yields the score matrix $S \in \mathbb{R}^{n\times k}$.

We then individually align the composites with $t_1$ (the vector of first-PC scores across observations). Because Pearson correlation is invariant to positive affine transformations, the choice between z-standardized and original composite scores does not affect $\rho$:
\[
\rho_{t_1,U} = \operatorname{corr}(t_1, U),
\qquad
\rho_{t_1,R} = \operatorname{corr}(t_1, R).
\]
These values capture how strongly the dominant principal component $t_1$ aligns with each composite considered separately. Because the utility and risk composites may themselves be correlated, separate correlations can overstate the overall alignment. We therefore also consider a joint model in which $t_1$ is regressed on both composites simultaneously 

\[
t_1 = \beta_0 + \beta_U U + \beta_R R + \varepsilon.
\]
The coefficient of determination $R^2$ from this regression expresses the total fraction of variance in the component scores that is explained jointly by the two composites. Thus, while $\rho_{t_1,U}$ and $\rho_{t_1,R}$ describe individual associations, $R^2$ provides a single summary of joint alignment. For interpretation we report three sets of results:
\begin{enumerate}
  \item the proportion of variance in the measures explained by the first
        principal component (from the PCA output);
  \item the individual correlations $\rho_{t_1,U}$ and $\rho_{t_1,R}$, together with
        their squared values $\rho_{t_1,U}^2$ and $\rho_{t_1,R}^2$, which indicate
        the proportion of variance in each composite accounted for by the
        component scores;
  \item the coefficient of determination $R^2$ from the joint regression, summarizing the total fraction of variance in PC1             explained jointly by both composites -- clarifying whether PC1 can be interpreted as a single risk--utility summary axis         or whether other patterns dominate.
\end{enumerate}

This approach has three caveats: it relies on linear correlation, the composites $U$ and $R$ are themselves averages of subsets of the variables that produced the principal components, so a high $R^2$ partly reflects this structural overlap rather than providing independent validation, and it assumes that the mean composites meaningfully capture the underlying risk and utility constructs. 
When alignment is weak or absent, additional components can be examined for secondary or dataset-specific patterns; if none show clear alignment, the PCA is interpreted descriptively via its loadings.

Applying this framework to our EU-SILC data, Table~\ref{tab:alignment_eusilc} shows that the first principal component (PC1) correlates strongly with both the utility composite ($\rho = 0.86$) and the risk composite ($\rho = 0.95$). The linear regression yielded $R^{2} = 0.97$, indicating that PC1 is almost
entirely explained by the two composites. This reflects the inherent risk-utility trade-off: methods that achieve higher utility tend to simultaneously exhibit higher risk, causing both constructs to align along the same dominant axis.

\begin{table}[ht]
\caption{Alignment of PC1 with utility and risk composites (EU-SILC)}
\label{tab:alignment_eusilc}
\centering
\begin{tabular}{lrrrrr}
  \hline
dataset & $\rho_{s,U_c}$ & $\rho_{s,R_c}$ & $\rho^2_{s,U_c}$ & $\rho^2_{s,R_c}$ & $R^2_{joint}$ \\
  \hline
EU-SILC & 0.86 & 0.95 & 0.74 & 0.89 & 0.97 \\
   \hline
\end{tabular}

\end{table}

\subsubsection{Blockwise PCA R-U Map} \label{sec:blockwise_pca}

 In the blockwise PCA approach, two separate principal component analyses are constructed: one on the set of disclosure risk measures and one on the set of utility measures \parencite[for utility measures only, see][]{2022_Dankar_PCA}. The first principal component of the utility block is then plotted on the $x$-axis and the PC1 of the risk block on the $y$-axis. Each point in the resulting two-dimensional plot represents an anonymization approach, while the two axes serve as composite indices summarizing the respective blocks. Unlike the simple means used for the composites in the composite R-U map in Section \ref{sec:theory_ru_map}, this PCA approach takes into account the correlation structure within each block and assigns weights to measures based on their contribution to overall variance. This allows for a more informed aggregation, where strongly varying and interrelated measures are emphasized, rather than treating all measures as equally important, while also reducing potential imbalance arising from unequal numbers of risk and utility measures. 
 
 However, this summarization comes at a cost: the axes are unitless and data-driven, and their interpretation depends on the specific loadings of the first principal components. In addition, relationships between disclosure risk and utility that are primarily associated with variation reflected in higher-order components may become less visible. For example, consider two utility measures capturing different aspects of data quality, such as predictive performance and logical consistency (e.g., absence of impossible household compositions or invalid demographic combinations). If PC1 in the utility PCA is mainly driven by variation in predictive performance, while logical consistency is primarily represented in higher-order components, then a relationship between disclosure risk and logical consistency may be less visible in the blockwise PCA visualization.
 
\begin{figure}[H]
    \centering
    \includegraphics[width=\linewidth]{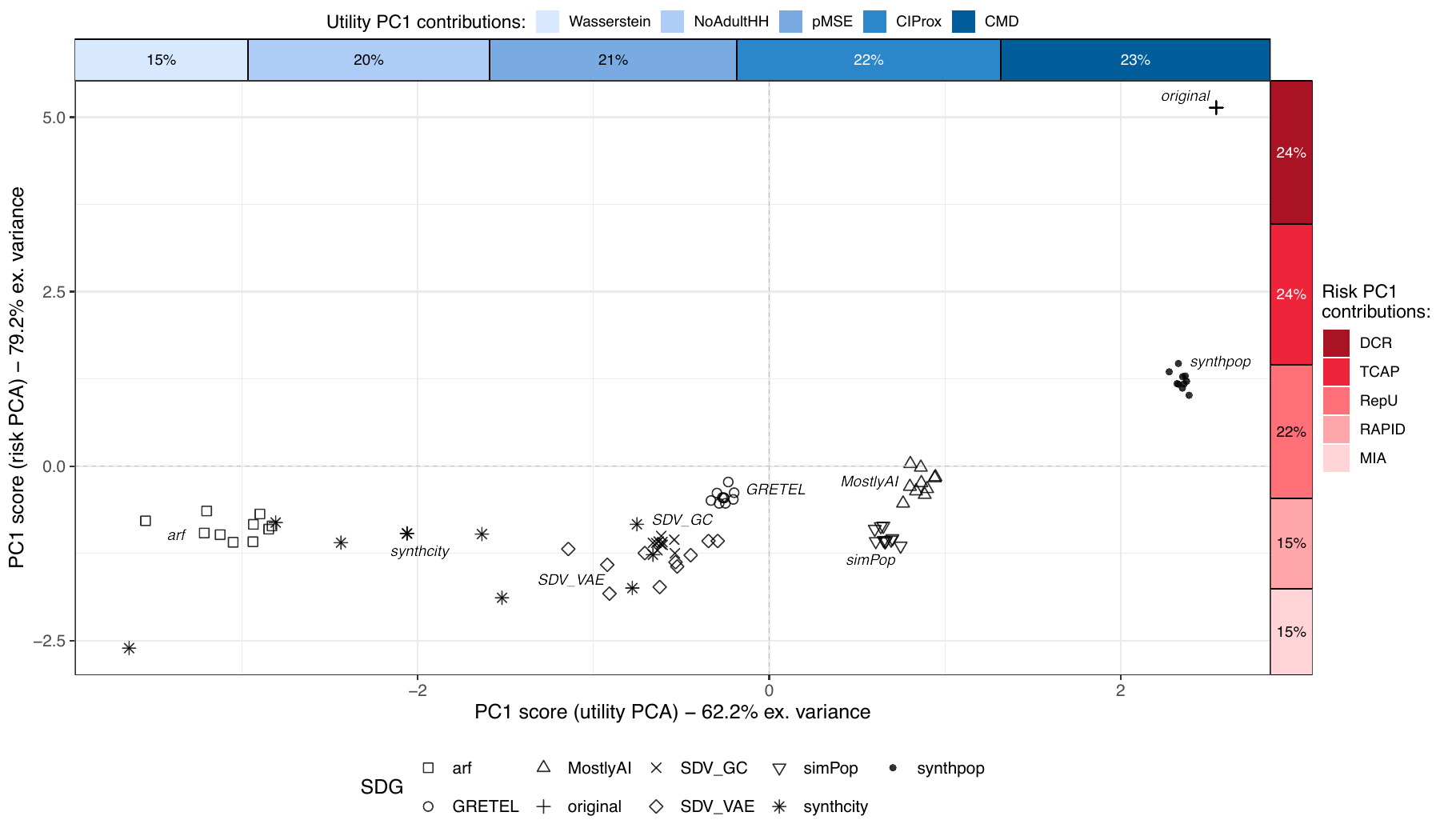}
    \caption{
    Blockwise PCA summary. X-axis: PC1 from utility measures; Y-axis: PC1 from risk measures (points = SDGs; blue = Pareto-optimal). Stacked bars show each measure's contribution to PC1, computed from squared loadings and normalized to 100\%    (labels shown for contributions $\geq$ 5\%). Edge colors indicate loading sign (black = positive, red = negative).
    }
    \label{fig:block_pca}
\end{figure}

To support interpretation despite these limitations, the squared normalized loadings of PC1 can be visualized as stacked bar charts aligned to each axis (Utility PC1 and Risk PC1), showing which measures contribute most to the composite scores (cf. Figure~\ref{fig:block_pca}). Optionally, Pareto-optimal anonymization approaches (identified in the original composite score space) can be highlighted, and, when comparing multiple datasets for each anonymization approach, group structures can be visualized using ellipses to indicate clustering or method families.

Figure~\ref{fig:block_pca} shows the blockwise PCA for our EU-SILC data. Since both Utility PC1 and Risk PC1 are linear combinations of their respective z-standardized measures, they act as a data-driven composite -- assigning weights based on the variance structure of the data rather than treating all measures equally -- and their joint scatter approximates the composite R-U map (cf. Figure~\ref{fig:composite_RU_extended}). The stacked bar charts in Figure \ref{fig:block_pca} display each measure's contribution to its corresponding PC1, calculated from squared loadings (representing variance contributions) and normalized to 100\%. The relatively uniform distribution of these contributions confirms that the measures contribute roughly equally to each principal component. Utility PC1 explains 62.2\% of variance across the five utility measures, while Risk PC1 explains 79.2\% across the five risk measures. While the blockwise PCA approach adds computational complexity and reduces interpretability compared to simple averaging, it provides empirical validation that the mean-based composite scores are defensible; the uniform loadings confirm that no single measure dominates and that simple averaging is not merely a convenience but an empirically justified choice. In scenarios with heterogeneous or redundant measures, this approach would reveal a structure that simple averaging obscures.

\section{Comparative Overview of Visualization Methods}\label{sec:method_overview}

In addition to the visualization tools described in Section~\ref{sec:viz_tools}, this section presents a structured comparison of these approaches based on an evaluation framework developed by the authors, drawing on criteria from the disclosure risk, data utility, and visualization literature cited throughout this paper. Similar capability-based comparison frameworks have been employed in the multi-objective optimization literature \parencite{Nagar_2023_Visualization}. The criteria capture analytical capabilities (e.g., identifying patterns, relationships, and anomalies), trade-off visualization properties (e.g., 
representation of Pareto-optimal solutions), and usability aspects such as interpretability and scalability with increasing numbers of methods or measures.

The evaluation is summarized in Table~\ref{tab:capability_comparison2}. No single method excels across all criteria, suggesting that combining multiple visualization approaches often provides the most complete understanding of risk--utility 
trade-offs. For a discussion and practical guidance, see Section~\ref{sec:discussion} and Section \ref{sec:conc_recomm}.

\begin{table}[ht]
\hspace{-1.3cm}
\scriptsize
\begin{tabular}{>{\raggedright\arraybackslash}p{8.2cm}cccccc}
\toprule
\textbf{Capability} &
\makecell{\textbf{Heat-}\\\textbf{maps}} &
\makecell{\textbf{Dot}\\\textbf{Plots}} &
\makecell{\textbf{Composite RU}\\\textbf{Scatterplots}} &
\textbf{PCP} &
\makecell{\textbf{Radial}\\\textbf{Profile Plots}} &
\makecell{\textbf{PCA }\\\textbf{Biplots}} \\
\midrule
\textbf{Analytical Capabilities}                                                              &&&&&& \\

Detecting systematic differences across methods                            & \checkmark  & \checkmark & \checkmark  & \checkmark & \checkmark & \checkmark   \\
Correlations between risk and utility measures                                                         & $\sim$   & $\sim$  & $\sim$  & $\sim$  & $\sim$  & \checkmark   \\
Uncertainty depiction (e.g. measurement or sampling error)                                                      & $\times$    & $\times$   & \checkmark  & $\times$   & $\times$   & \checkmark   \\
Outlier detection                                                          & $\sim$   & $\sim$  & $\sim$  & $\sim$  & $\sim$  & \checkmark   \\

\midrule
\textbf{Trade-off Visualization}                                                              &&&&&& \\

Displaying of Pareto-optimal methods                                       & \checkmark  & \checkmark & \checkmark  & \checkmark & \checkmark &  $\sim$   \\
Displaying Pareto-Front                                                    & $\times$    & $\times$   & \checkmark  & $\times$   & $\times$   & $\times$   \\
Support of acceptable thresholds                                           & \checkmark  & \checkmark & $\sim$  & \checkmark & $\sim$  & $\sim$  \\
\midrule
\textbf{Usability \& Scalability}                                                             &&&&&& \\

Scalability with number of methods                                         & \checkmark  & $\sim$  & \checkmark  & \checkmark & $\times$   & \checkmark   \\
Scalability with number of risk and utility measures                                         & \checkmark  & $\sim$  & \checkmark  & \checkmark & $\times$   & \checkmark   \\
Comparison of methods across multiple datasets                                   & $\sim$   & $\times$   & \checkmark  & $\times$   & $\times$   & \checkmark   \\
Intuitive interpretation for non-technical stakeholders                    & \checkmark  & \checkmark & \checkmark  & \checkmark & \checkmark & $\sim$    \\
Raw value of each measure can be displayed                                                       & \checkmark   & \checkmark  & $\times$  & \checkmark & \checkmark  & $\times$   \\
\bottomrule
\end{tabular}
\caption{Capability comparison across visualization methods, organized
by analytical capabilities, trade-off visualization, and usability
(\checkmark = good, $\times$ = poor/unsupported, $\sim$ = partial/mixed).  Assessments reflect the
authors' qualitative evaluation based on the illustrative applications presented
in this paper, the visualization literature cited throughout, and theoretical
properties of the methods; they are not derived from a formal user study or
empirical testing.}
\label{tab:capability_comparison2}
\end{table}

\section{Discussion}\label{sec:discussion}
Building on the visualization approaches presented in Section~\ref{sec:viz_tools} and the structured comparison in Table~\ref{tab:capability_comparison2}, we now discuss practical guidance for method selection. For initial screening and detailed inspection of individual measures, heatmaps and dot plots provide complementary strengths. Heatmaps are especially well suited for initial diagnostics and for communicating metric-level performance in a concise and structured way. The choice of layout depends on the primary comparison task: methods-as-rows facilitates scanning the full profile of a single method across measures, while methods-as-columns facilitates comparing multiple methods on a single measure. The encoded information is identical in both orientations. 

Dot plots can enhance heatmap analysis when visualizing variation across risk and utility dimensions is needed, since the spread of points is easier to interpret visually than color intensity and numbers alone. Dot plots further excel in showing individual measure values with minimal chart junk \parencite{Tufte_1983_Visual} and can easily be expanded with distributional information using boxplots (see Figure~\ref{fig:dotplot}), for instance when multiple datasets are available and uncertainty should be depicted. However, dot plots may suffer from overplotting when too many methods or measures are displayed; in such cases, jittering or small multiples can still preserve readability.

When using composite scores for Pareto identification, reliability depends on whether measures within each block show reasonably consistent values for each method. A method with highly disparate measure values may appear Pareto-optimal based on its mean but has critical vulnerabilities the composite obscures (because of outliers). Heatmaps reveal within-method disparities through mixed colors within rows, while dot plots clearly show the spread of individual measure values. As a diagnostic check, one can inspect the range or standard deviation of measures within each block, or verify that PC1 of a blockwise PCA explains a high proportion of variance with relatively uniform loadings. When substantial disparities exist, measure-specific examination provides necessary context beyond composite analysis.

Composite scatterplots -- whether based on simple averages or blockwise PCA -- provide an intuitive risk-utility visualization that is accessible to both technical and non-technical audiences. The main advantage of using blockwise PCA over simple averaging is empirical validation: if PC1 in each block explains a high proportion of variance (e.g., >70\%) and loadings are relatively uniform, this justifies the dimensionality reduction. When loadings are heterogeneous, blockwise PCA reveals which measures dominate each composite, information that simple averaging can obscure. However, the number of observations -- i.e., the number of anonymization approaches -- is typically small in practice, which can limit the reliability of internal consistency measures such as Cronbach's $\alpha$ and McDonald's $\omega$ for composite scores. For this reason, we report these coefficients when composite scores are used but advise to interpret them with caution. In low-n settings, a more stable strategy may be to rely on conceptual grouping of measures rather than on formal internal consistency statistics alone.

For multivariate profile visualization, radial profile charts such as origami plots \parencite{Duan_2023_Origami} and parallel coordinate plots both deliver a gestalt representation of the full performance profile, but differ in scalability and accessibility. Origami plots are visually memorable and intuitive for stakeholder communication, but the number of approaches and measures should be limited as overlapping polygons become difficult to distinguish; small multiples are a viable option when more approaches need to be shown. PCPs offer a more scalable alternative, preserving readability and pattern detection at higher dimensions.

PCA-based biplots not only visualize method performance but also reveal the relationships among measures themselves. While the biplot visualization facilitates the simultaneous comparison of multiple risk and utility measures, its interpretive value depends on the extent to which the first two principal components capture the overall variance in the data. If these two dimensions explain only a small proportion of the variance, the plot may provide an incomplete view of the risk and utility properties of the anonymization strategies. In such cases, examining additional components or using alternative dimensionality reduction techniques may be necessary. 

Biplots are only applicable when all anonymization approaches under comparison use the same set of risk and utility measures. This becomes problematic when the disclosure risk of a non-perturbative anonymization approach is evaluated using metrics such as, e.g., the number of cases violating $k$-anonymity \parencite{2002_Sweeney}, which are not applicable to perturbative anonymization approaches that, e.g., swap values of quasi-identifiers. Because the set of relevant metrics may not overlap, direct comparison in a single biplot is not appropriate in such scenarios.

In the right use cases, such as for comparing the performance of various SDGs, biplots can provide clear insights into the trade-off between disclosure risk and data utility. In this context, the biplot not only highlights overall performance but also reveals how risk and utility measures relate to each other, supporting a more nuanced understanding of anonymization quality. To enhance biplot interpretability, a sign convention can be established to ensure that the first principal component always correlates positively with utility measures (as briefly discussed in Section~\ref{sec:theory_pca}). This convention makes the visualization more intuitive by ensuring that utility measure loadings consistently point toward the positive PC1 direction, so approaches on the right side of the biplot tend to perform better on utility measures. The vertical axis and the precise positioning of approaches still depend on the full covariance structure of all measures.
Without such a convention, the arbitrary sign of principal components can lead to confusion when comparing across datasets or analyses. 

While Pareto-optimality can be identified in the original composite score space and subsequently highlighted in the biplot, it should be noted that Pareto-optimality is not geometrically preserved under PCA projection. A method that is Pareto-optimal in the original risk–utility space need not occupy an extremal position in the biplot, as the PCA axes are linear combinations of all measures rather than pure risk or utility dimensions. Pareto highlighting in biplots should therefore be interpreted cautiously and always with reference to the original composite scores.

We close with a broader conceptual reflection on the risk–utility framing underlying this paper. Although risk and utility are often framed as a trade-off, as we also did in this paper, synthetic data can achieve comparable or even slightly improved utility relative to the original dataset, particularly when evaluated via downstream predictive tasks \parencite{Pilgram_2025_Magnitude}. When utility exceeds that of the original data while risk remains the same or decreases, the framework naturally identifies these as Pareto-optimal solutions, illustrating that it captures both antagonistic and synergistic relationships between risk and utility.

\subsection{Extensions and Alternative Approaches}

For specialized use cases, bivariate color scales \parencite{vonMayr_1874_Gutachten, Wainer_1980_Empirical} can encode two variables simultaneously -- for instance, displaying both risk and utility dimensions through color intensity and hue in geographic or network visualizations.

We considered but did not pursue several alternative approaches. Nonlinear dimensionality reduction methods like t-SNE  \parencite{vanderMaaten_2008_Visualizing} or UMAP \parencite{McInnes_2020_UMAP} could capture complex relationships but lack PCA's interpretable loadings. The interpretable self-organizing map \parencite[iSOM;][]{Nagar_2023_Visualization} shows promise for dense Pareto-optimal sets with many candidates. However, in anonymization contexts where a limited number of approaches are typically compared, SOM grids become under-determined and highly sensitive to hyperparameters. The visualization approaches presented here are better suited to small-sample scenarios common in practice.

Beyond the methods explored here, several directions warrant further investigation. Alternative composite construction methods, such as median-based or stakeholder-weighted composites, could improve the robustness of Pareto identification. Relatedly, methods for identifying Pareto-optimal approaches under uncertainty warrant further investigation, as composite risk and utility scores are estimated from multiple measures or repeated syntheses and may therefore introduce variability in Pareto classification. Future work could explore approaches that explicitly account for this uncertainty, for example through bootstrap-based Pareto fronts or probabilistic dominance criteria. More broadly, developing comparison frameworks that accommodate incomplete measure coverage would enhance practical applicability. A formal user evaluation involving both technical and non-technical stakeholders could further validate and refine the proposed comparison framework. Finally, interactive preference elicitation methods that guide decision-makers toward their optimal risk--utility trade-off -- for instance by presenting hypothetical Pareto fronts and learning user preferences through iterative selections \parencite{Yang_2025_Interactive} -- offer a promising avenue beyond post-hoc visual comparison of static trade-offs.

\section{Conclusion}\label{sec:conclusion}
\subsection{Summary of Contributions}

This paper addresses anonymization approach selection as a genuine multivariate optimization problem. Traditional risk-utility visualizations typically compare a single risk measure against a single utility measure, though multiple measures are often calculated for each dimension. We present and systematically compare six visualization approaches for simultaneous evaluation of multiple risk and utility measures: heatmaps, dot plots, composite scatterplots, parallel coordinate plots, radial profile charts, and PCA-based biplots. Through systematic Pareto-optimal approach identification applied across all approaches, we demonstrate that simultaneously visualizing multiple measures provides richer evaluation than selecting single representatives. Our comparative analysis reveals that visualization choice should align with analytical objectives: PCA approaches excel at revealing measure relationships and multivariate structure, while simpler approaches facilitate initial screening or intuitive assessment.

\subsection{Key Recommendations}\label{sec:conc_recomm}

Effective anonymization approach selection requires integrating technical risk-utility assessment with broader organizational considerations including legal frameworks, data sensitivity, and institutional risk tolerance \parencite{2017_Templ, 2012_Hundepool}. For the technical assessment component, we recommend employing multiple complementary visualizations tailored to analytical objectives. We recommend first analyzing the risk and utility measures univariately, and then complementing this view with a multivariate perspective using the visualization methods described in this article. 
For initial screening, heatmaps efficiently reveal overall performance patterns while dot plots and their distributional extensions clearly display individual measure values across risk and utility dimensions. 
Before compositing, internal consistency of each block can be checked 
(McDonald's $\omega$, as a heuristic) to verify that averaging across 
measures is justified. A composite R--U scatterplot can then provide an intuitive trade-off summary; overlaying a risk-tolerance threshold and identifying the Pareto-optimal set and knee point supports systematic selection. For deeper structural insight, PCA-based biplots reveal measure relationships and multivariate structure, and parallel coordinate plots effectively display high-dimensional profiles. Pareto-optimal approaches can be identified and highlighted consistently across these visualization types. However, composite-based identification of Pareto-optimal approaches requires checking whether aggregation adequately represents each approach's performance across measures.

\section*{Acknowledgment \& Disclosure}

\subsection*{Acknowledgment}
This work was funded by the Swiss National Science Foundation (SNSF) with grant ``Harnessing event and longitudinal data in industry and health sector through privacy preserving technologies'' (Grant Number \href{https://data.snf.ch/grants/grant/211751}{211751}).

\subsection*{Disclosure of Interests}
The authors have no competing interests to declare that are relevant to the content of this article.

\newpage
\appendix
\section{Technical Specifications}\label{app:SDGs}

\begin{table}[htbp]
\centering
\caption{Overview of the used synthetic data generators}
\label{tab:SDGs}
\normalsize
\begin{adjustbox}{width=\linewidth}
\begin{tabular}{l  p{6cm} l l l}
\toprule
\textbf{Name} &
\textbf{Method(s)} & \textbf{Software} & \textbf{Authors} & \textbf{Version} \\
\midrule
synthpop
  & CART
  & R package
  & \cite{Nowok_2016_Synthpop} & 1.9.1 \\
Synthetic Data Vault
  & GaussianCopula; VAE 
  & Python package
  & \cite{Patki_2016_Synthetic} & 1.18.0 \\
simPop
  & Multinomial log-linear models; random draws; random forest (alt.)
  & R package
  & \cite{Templ_2017_Simulation} & 2.1.3 \\
Mostly AI
  & Transformers; GANs; VAEs; autoregressive networks
  & Mostly AI (AT)
  & \cite{MostlyAI_2025_Mostly} & 4.2.3 \\
Gretel
  & Synthetic ACTGAN
  & Gretel Labs (US)
  & \cite{GretelAI_2025_Gretel} & 0.22.16 \\
 arf
  & Adversarial random forests
  & R package
  & \cite{Watson_2023_Adversarial} & 0.2.0 \\
synthcity
  & Tabular GAN
  & Python package
  & \cite{Qian_2023_Synthcity} & 0.2.11 \\
\bottomrule
\end{tabular}
\end{adjustbox}
\end{table}

\begin{table}[htbp]
\centering
\caption{Specifications of the risk and utility measures}
\label{tab:measure_specification}
\normalsize
\begin{adjustbox}{width=\textwidth}
\begin{tabular}{l >{\raggedright\arraybackslash}p{5cm} p{11cm}}
\toprule
\textbf{Type} & \textbf{Measure \& Abbreviation} &  \textbf{Specification} \\
\midrule

Risk    & Replicated Uniques (\texttt{RepU} )       & Key variables: \texttt{age}, \texttt{db040}, \texttt{rb090}, \texttt{hsize}, and \texttt{pb190} \\
Risk    & Distance to Closest Record (\texttt{DCR} )       & Metric: ratio of mean distances train/holdout \newline Matching variables: all available variables \newline Holdout-set size: 25\% \\
Risk    & Membership Inference Attack (\texttt{MIA})      &   Model type: random forest \newline Predictor variables: all available variables \newline Holdout-set size: 25\% \\
Risk    & Targeted Correct Attribution Probability (\texttt{TCAP} ) &
Key variables: \texttt{age}, \texttt{db040}, \texttt{rb090}, \texttt{hsize}, and \texttt{pb190} \newline Target variable: \texttt{pgrossIncome} \\
Risk    & Risk of Attribute Prediction-Induced Disclosure (\texttt{RAPID})           & Model type: random forest \newline Key variables: \texttt{age}, \texttt{db040}, \texttt{rb090}, \texttt{hsize}, and \texttt{pb190} \newline Target variable: \texttt{pgrossIncome} \newline Threshold: $\epsilon = 0.05$\\
\midrule
Utility & Confidence Interval Overlap (\texttt{CIProx})                                                         &  Target variable: \texttt{pgrossIncome} \newline CI-level: 95\% \\
Utility & Propensity Mean Squared Error (\texttt{pMSE} ) &
Model type: random forest \newline Predictor variables: \texttt{db040}, \texttt{hsize}, \texttt{pb220a}, \texttt{rb090}, \texttt{pl031}, and \texttt{pgrossIncome} \\
Utility & Wasserstein Distance (\texttt{Wasserstein})                        & Mean Wasserstein-1 distance across all numeric variables of the EU-SILC dataset, each normalized by its own IQR, between original and synthetic distributions.\\
Utility & Households with only children (\texttt{NoAdultHH})    & Count of households, identified by \texttt{db030}, in which all members are under 18 years of age; should be zero in valid synthetic data.  \\
Utility & Correlation Matrices Differences (\texttt{CMD})                                          & Correlation variables: \texttt{pgrossIncome}, \texttt{hy140g}, \texttt{hx050}, and \texttt{age}  \newline Method: spearman\\
\bottomrule
\end{tabular}
\end{adjustbox}
\end{table}

\begin{tablenotes}
\tiny
\item \textit{Note}: Variable names follow EU-SILC coding.
\texttt{age} = age of the person,
\texttt{db040} = country code,
\texttt{rb090} = gender,
\texttt{hsize} = household size,\\
\texttt{hx050} = equivalized household size,
\texttt{pb190} = marital status,
\texttt{pgrossIncome} = gross personal income,
\texttt{pl031} = employment status, \\
\texttt{hy140g} = household taxes and social contributions ,
\texttt{db30} = household id
\end{tablenotes}

\begin{table}[H]
\centering
\scriptsize
\caption{SDG-specific settings used for the synthesis of the EU-SILC dataset. Each SDG
generated $m = 10$ synthetic datasets of size $n = 13\,513$. Seeds were set to $1$--$10$
per iteration unless otherwise noted. Parameters not listed were kept at the respective
package defaults.}
\label{tab:sdg_settings_eusilc}
\vspace{0.25cm}
\setlength{\tabcolsep}{3pt}
\renewcommand{\arraystretch}{0.85}
\begin{tabular}{p{2.8cm} p{5.2cm} p{6.0cm}}
  \hline
  \textbf{SDG} & \textbf{Parameter} & \textbf{Specification} \\
  \hline
  \texttt{synthpop}
    & Synthesis method           & CART (default) for all variables \\
    & Synthesis order            & Sequential (column order, default) \\
    & Semi-continuous treatment  & \texttt{pgrossIncome} \\
    & Seeds                      & 1--10 (one per iteration) \\[1pt]
  \texttt{arf}
    & Pipeline                   & \texttt{adversarial\_rf()} $\to$ \texttt{forde()} $\to$ \texttt{forge()} \\
    & Seed                       & \texttt{set.seed(123)}; all other parameters at defaults \\[1pt]
  \texttt{simPop}
    & Seed                       & \texttt{set.seed(123)}, fixed for all iterations \\
    & Input specification        & \texttt{hhid = db030; hhsize = hsize; strata = db040;} weight = \texttt{rb050} (normalized: $/616.493$) \\
    & Structure                  & \texttt{method = "direct";} \texttt{basicHHvars}: \texttt{age, rb090, db040, hx050} \\
    & Categorical variables      & \texttt{method = "multinom";} \texttt{additional}: \texttt{pl031, pb220a, pb190, pe040, pl111} \\
    & Continuous: \texttt{pgrossIncome} & \texttt{method = "multinom"; upper = 200\,000; equidist = FALSE; log = TRUE;} \texttt{regModel}: $\sim$\texttt{rb090 + hsize + pl031 + pb220a + pb190 + pe040 + pl111} \\
    & Continuous: \texttt{hy140g} & \texttt{method = "multinom"; upper = 200\,000; equidist = FALSE; log = TRUE} \\
    & Income components          & \texttt{pgrossIncome} decomposed into \texttt{py010g}--\texttt{py140g} (10 components); conditional on \texttt{pl031, pb220a, rb090, pe040, pb190, pl111}; \texttt{replaceEmpty = c("sequential", "min")} \\
    \\[1pt]
  \texttt{SDV} \newline(GaussianCopula)
    & Metadata                   & Column types defined in \texttt{metadata\_eusilc.json} (categorical, numerical, and id fields as detected by SDV and manually verified) \\
    & Sampling batch size        & 1\,000 \\
    & Reproducibility note       & Model re-initialized and re-fitted per iteration to avoid identical samples (known SDV issue) \\[1pt]
  \texttt{SDV} (TVAE)
    & Epochs                     & 1\,000 \\
    & \texttt{enforce\_min\_max\_values} & \texttt{True} \\
    & \texttt{enforce\_rounding} & \texttt{False} \\
    & Sampling batch size        & 1\,000 \\
    & Reproducibility note       & Model re-initialized and re-fitted per iteration to avoid identical samples (known SDV issue) \\[1pt]
  \texttt{synthcity} \newline(TabularGAN)
    & \texttt{n\_units\_latent}  & 128 \\
    & \texttt{batch\_size}       & 1\,000 \\
    & \texttt{n\_iter}           & 1\,000 (default training iterations) \\
    & Seeds                      & \texttt{random\_seed = i}, $i = 1$--$10$ \\[1pt]
  Mostly AI
    & Interface                  & Web platform (\texttt{app.mostly.ai}) \\
    & Max training epochs        & 1\,000 \\
    & Value protection           & Off \\[1pt]
  Gretel (ACTGAN)
    & Interface                  & Web platform; config via \texttt{.yml} \\
    & Epochs                     & 1\,000 \\
    & Generator dimensions       & $[1024, 1024]$ \\
    & Discriminator dimensions   & $[1024, 1024]$ \\
    & Generator learning rate    & 0.0001 \\
    & Discriminator learning rate & 0.00033 \\
    & \texttt{batch\_size}       & 1\,000 \\
    & Privacy filters            & \texttt{outliers = auto; similarity = auto} \\[1pt]
  \hline
\end{tabular}
\smallskip
\begin{minipage}{\textwidth}
\tiny
\textit{Note}: Variable names follow EU-SILC coding.
\texttt{age} = age of the person;
\texttt{db030} = household id;
\texttt{db040} = country code;
\texttt{hsize} = household size;
\texttt{hx050} = equivalized household size;
\texttt{rb050} = survey weight;
\texttt{rb090} = gender;
\texttt{pb190} = marital status;
\texttt{pb220a} = citizenship;
\texttt{pe040} = highest education level (ISCED);
\texttt{pl031} = employment status;
\texttt{pl111} = months in full-time employment;
\texttt{pgrossIncome} = gross personal income;
\texttt{py010g}--\texttt{py140g} = income components of \texttt{pgrossIncome};
\texttt{hy140g} = household taxes and social contributions.
\end{minipage}
\end{table}

\section{Numerical Illustration: Scaling Effects on Averaged Composites and Pareto Identification}
\label{app:pareto}

Table~\ref{tab:scaling_example} illustrates how the choice of scaling method affects composite scores and consequently the identified Pareto-optimal set using four constructed SDGs with two utility measures ($u_1, u_2$) and two risk measures ($r_1, r_2$). Raw values are: SDG1 $(u_1{=}0.86,\, u_2{=}62.2,\, r_1{=}0.89,\, 
r_2{=}152.2)$; SDG2 $(0.62, 65.0, 0.47, 112.3)$; SDG3 $(0.33, 145.9, 0.12, 154.2)$; SDG4 $(0.06, 127.5, 0.71, 98.8)$. The rank columns already reveal differences across scaling methods.

Approach $i$ is Pareto-optimal if no $j \neq i$ exists with $\bar{u}_j \geq \bar{u}_i$ and $\bar{r}_j \leq \bar{r}_i$ with at least one strict inequality. The Pareto set differs across scaling methods because dominance is evaluated on composite scores rather than on the raw measure vectors. This is a deliberate choice: composite-based Pareto identification is consistent with the two-dimensional visualizations presented in this paper and improves interpretability for practitioners, at the cost of making the result sensitive to the scaling method -- a limitation that should be acknowledged when reporting results.

\begin{table}[H]
\centering
\caption{Effect of scaling on composite scores and Pareto identification.
$\bar{u}$ and $\bar{r}$ denote mean utility and risk composites;
$\text{rk}_u$ = utility rank (descending) and $\text{rk}_r$ = risk rank (ascending);
P~=~Pareto-optimal (\checkmark) or dominated (\texttimes).}
\label{tab:scaling_example}
\small
\setlength{\tabcolsep}{4pt}
\begin{tabular}{l rrrrr rrrrr rrrrr}
\toprule
& \multicolumn{5}{c}{\textbf{Raw (unscaled)}}
& \multicolumn{5}{c}{\textbf{Min--max scaled}}
& \multicolumn{5}{c}{\textbf{Z-score standardized}} \\
\cmidrule(lr){2-6} \cmidrule(lr){7-11} \cmidrule(lr){12-16}
\textbf{SDG}
& $\bar{u}^{\text{raw}}$ & $\bar{r}^{\text{raw}}$ & $\text{rk}_u$ & $\text{rk}_r$ & P
& $\bar{u}^{\text{mm}}$ & $\bar{r}^{\text{mm}}$ & $\text{rk}_u$ & $\text{rk}_r$ & P
& $\bar{u}^{\text{z}}$ & $\bar{r}^{\text{z}}$ & $\text{rk}_u$ & $\text{rk}_r$ & P \\
\midrule
SDG1 & 31.53 & 76.54 & 4 & 3 & \texttimes & 0.500 & 0.982 & 2 & 4 & \texttimes & \phantom{$-$}0.141 & \phantom{$-$}1.063 & 2 & 4 & \texttimes \\
SDG2 & 32.81 & 56.38 & 3 & 2 & \texttimes & 0.367 & 0.349 & 4 & 1 & \checkmark & $-0.220$ & $-0.486$ & 3 & 1 & \checkmark \\
SDG3 & 73.12 & 77.16 & 1 & 4 & \checkmark & 0.669 & 0.500 & 1 & 3 & \checkmark & \phantom{$-$}0.388 & $-0.231$ & 1 & 3 & \checkmark \\
SDG4 & 63.78 & 49.75 & 2 & 1 & \checkmark & 0.390 & 0.383 & 3 & 2 & \checkmark & $-0.309$ & $-0.347$ & 4 & 2 & \texttimes \\
\bottomrule
\end{tabular}
\end{table}

\section{Robust PCA}\label{app:robPCA}
\begin{figure}[H]
  \centering
  \includegraphics[width=\linewidth]{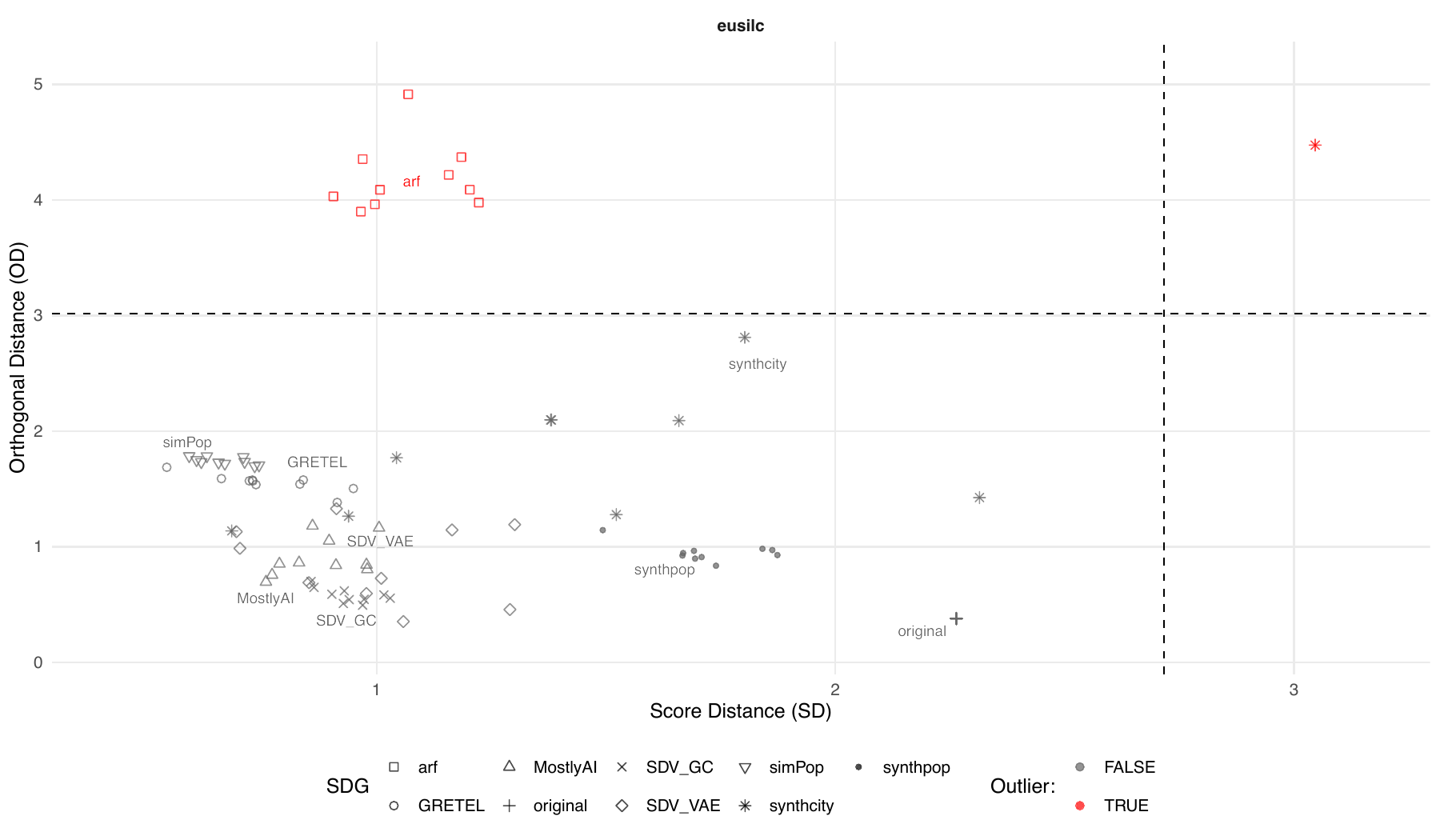}
  \caption{Robust PCA diagnostics (“outlier map”) for the data shown in Figure~\ref{fig:joint_pca}. The x-axis shows the Score Distance (SD) in the retained PC space (spanned by \(k = 2\) components), and the y-axis shows the Orthogonal Distance (OD) to that space. Each point is an SDG; color indicates the robust outlier flag. Large SD indicates leverage within the PC space; large OD indicates poor reconstruction (the observation lies far outside the subspace). Cutoffs follow \textcite[p.~66]{Hubert_2005_ROBPCA}: SD is compared to \(\sqrt{\chi^2_{k,0.975}}\) and OD to a cutoff derived via Wilson--Hilferty approximation.
  }
  \label{fig:robust_diagnostics_pca}
\end{figure}

To screen for outliers, the score–distance / orthogonal-distance (SD-OD) diagnostic plot \parencite{Hubert_2005_ROBPCA} can be used. 
Following \textcite[p.~66]{Hubert_2005_ROBPCA}, the robust score distance and orthogonal distance are defined as
\[
\mathrm{SD}_i=\sqrt{\sum_{j=1}^k \frac{t_{ij}^2}{\ell_j}}, \qquad
\mathrm{OD}_i=\big\|x_i-\mu - P_{p,k}\, t_i\big\|_2,
\]
where $t_{ij}$ are (robust) scores on PC $j$, $\ell_j$ the corresponding eigenvalues, $\mu$ the robust center, and $P_{p,k}$ the loading matrix which all need to be estimated. 
Intuitively, SD measures leverage within the $k$-dimensional PC subspace, i.e. the space spanned by the first $k$ components retained in the model. 
It is proportional to the Mahalanobis distance of the score vector in this reduced space. 
Large SD means the observation has unusually large scores on one or more principal components -- i.e., it lies far from the center within the PC subspace. 
OD measures residual distance orthogonal to that subspace: it is the Euclidean distance between the observation and its projection onto  the $k$-dimensional PCA space. So a large OD means the point is poorly represented by the first $k$ PCs. \begin{table}[htbp]
\centering
\caption{SD--OD regions and their interpretation.}
\label{tab:sd-od}
\begin{tabular}{l p{7cm} p{5cm}}
\toprule
\textbf{Region (SD, OD)} & \textbf{Meaning} & \textbf{Typical follow-up} \\
\midrule
SD low, OD low  & Regular / typical; near the center of the data cloud and well represented by the first $k$ PCs. & No concern; representative SDG. \\
SD high, OD low & Good Leverage outlier; far within the PC subspace (large scores on one or more PCs) yet small reconstruction error. & Check influence; may be a valid extreme SDG. \\
SD low, OD high & Orthogonal outlier; not far in the PC plane but poorly reconstructed by the first $k$ PCs -- structure outside the captured subspace. & Inspect variables not explained by PCs; consider increasing $k$ or data issues. \\
SD high, OD high & Bad leverage; both far in the PC plane and poorly represented -- extreme and structurally unusual. & Strong candidate for anomaly; scrutinize or exclude. \\
\bottomrule
\end{tabular}
\end{table} For a quick look, the same diagnostic plot can be made with classical PCA; for outlier detection and stable cutoffs we prefer robust PCA, comparing SD to $\sqrt{\chi^2_{k, .975}}$ and the OD to a chi-squared-based cutoff derived from a Wilson--Hilferty normal approximation applied to $\mathrm{OD}^{2/3}$ \parencite[p.~66]{Hubert_2005_ROBPCA}. Points beyond either cutoff are flagged as outliers; cf.\ Table \ref{tab:sd-od}. Figure \ref{fig:robust_diagnostics_pca} shows the diagnostic plot of a robust version of the PCA of risk and utility measures. This diagnostic plot can provide valuable insights into which anonymization approaches exhibit distinctive performance.

\newpage
\printbibliography

\end{document}